\documentclass[final,5p,times,twocolumn,unknownkeysallowed]{elsarticle}

\usepackage{graphicx}
\usepackage{amssymb}
\usepackage{amsmath}
\usepackage{amsthm}
\usepackage{lineno}
\usepackage{hyperref}
\usepackage{multirow}
\usepackage{bigints}

\hypersetup{colorlinks=true,linkcolor=cyan,citecolor=cyan,urlcolor=blue,bookmarks=true}
\modulolinenumbers[200]

\biboptions{comma,sort&compress}

\journal{Journal Name}

\begin{document}

\begin{frontmatter}

\title{Gravitational lensing in 4-D Einstein-Gauss-Bonnet gravity in the presence of plasma}% Force line breaks with \\

\author[mainaddress1]
{Gulmina~Zaman~Babar}
\ead{gulminazamanbabar@yahoo.com}
\author[mainaddress2,mainaddress3,mainaddress4]
{Farruh~Atamurotov\corref{cor2}}\cortext[cor2]{Corresponding author}
\ead{atamurotov@yahoo.com and f.atamurotov@inha.uz}
\author[mainaddress5]
{Abdullah~Zaman~Babar}
\ead{abdullahzamanbabar@yahoo.com}

\address[mainaddress1]{School of Natural Sciences, National University of Sciences and Technology, Sector H-12, Islamabad, Pakistan}
\address[mainaddress2]{Inha University in Tashkent, Ziyolilar 9, Tashkent, 100170, Uzbekistan}
\address[mainaddress3]{Akfa University, Kichik Halqa Yuli Street 17,  Tashkent 100095, Uzbekistan}
\address[mainaddress4]{Ulugh Beg Astronomical Institute, Astronomy St. 33, Tashkent 100052, Uzbekistan}
\address[mainaddress5]{Department of Electrical Engineering, Air University, Islamabad, Pakistan}

\date{Received: date / Accepted: date}

\begin{abstract}
In this paper we have assumed a weak-field regime to explore the gravitational lensed photons
in a 4 dimensional Einstein-Gauss-Bonnet gravity, which is very much in the limelight these days. The investigation is conducted
in three distinct paradigms: uniform plasma, singular isothermal sphere and a non-singular isothermal sphere.
The lensing angle associated with the distribution factor of the medium is individually derived for each case and further
utilized to study the magnification of the image source brightness, selectively for the uniform plasma and singular isothermal sphere.
The attained results are brought forth in contrast with the standard Schwarzschild geometry.
\end{abstract}


\end{frontmatter}

\linenumbers

\section{Introduction} \label{intro}
As a matter of fact, the gravitational field of galaxy clusters and black holes is generally known to possess the ability to lens a nearby
passing light emitted from a source. The process takes place when the gravitational perturbations of the dark matter as well as the black holes
distort and magnify the incoming photons in its domain. This particular phenomenon
has not only helped the astronomers to discover the existence of invisible black holes but also
perceived the geometrical structure of the Universe. Refs \cite{Synge:1960b,Schnei:1999a,Perlick:2000a,Perlick:2004a}
render specific details regarding the fundamental tools required to study the lensing mechanism in the black hole vicinity.

Virbhadra and Ellis rekindled the lensing theory by considering a strong-field regime for the classical Schwarzschild
geometry and naked singularities \cite{Virbha:2000a,Virbha:2002a}. Bozza et al. based on the
Virbhadra-Ellis lens model developed an analytical approach to highlight the process \cite{Bozza:2001a}, afterwards Bozza
independently extended the investigation for the Schwarzschild, Reissner-Nordstr$\mathrm{\ddot{o}}$m and Janis-Newman-Winicour spacetimes
in the following paper \cite{Bozza:2002b}. The spinning Kerr black hole is explored by the authors in \cite{Bozza:2003a,Sevb:2004a,Bozza:2005a,Bozza:2006a}
to reveal its lensing properties. Some of the noticeable articles \cite{Eiroa:2002b,Eiroa:2004a,Eiroa:2005a,Virbha:2009b,Wei:2012b,Sotani:2015a,Zhao:2017a,Chak:2017a,Jin:2020a}
manifest an elaborate analysis in regard to the above referred field by considering the Reissner-Nordstr$\mathrm{\ddot{o}}$m, Braneworld,
Kerr-Taub-NUT, Born-infeld and Lee–Wick gravities.

The latter analysis has equally remained the subject of interest for a weak-field scenario. Bisnovatyi-Kogan and Tsupko with a great finesse
laid the groundwork for a weak-field lensing in the Schwarzchild  plasma \cite{Bin:2010a} and thereupon continued
their research for a variety of plasma mediums \cite{Tsp:2011a,Tsp:2014a,Tsp:2015a,Bis:2017a}. A rotating massive and compact objects,
non-Schwarzschild geometry, Bardeen, Hayward, braneworld, boosted Kerr
and Kerr-Newman gravities are addressed accordingly in \cite{Abu:2013a,Chak:2018a,Turi:2019a,Far:2016a,Abu:2017aa,Abu:2017a,Car:2018a,Babar:2020a,Chow:2020a,Far:2021a}.

In this paper our prime focus is to re-establish the process of weak-field gravitational lensing by taking into account the articles \cite{Bin:2010a,Abu:2013a,Abu:2017a}, in a plasma background of 4D Einstein-Gauss-Bonnet gravity (4D-EGB) \cite{Glav:2020a}, where D refers to the dimension of the space-time. Boulware et al. acquired a spherically symmetric static solution of EGB gravity in higher dimensions \cite{Boul:1985a}
by coupling the Einstein field action with the Gauss-Bonnet combination \cite{Bart:1985b}.
A distinctive analysis has been carried out in \cite{Cai:2002a} to identify the effects of thermodynamics in higher dimensional EGB Anti–de Sitter space-time. As of late, Glavan and Lin devised a modified gravity of Gauss-Bonnet theory in four dimensions by ignoring the Lovelock’s theorem
\cite{Dav:1971a}, which defines the theory of gravity in 4D only if the cosmological constant is provided, but rather made it
possible by reformulating the Gauss-Bonnet coupling constant $\alpha\rightarrow \alpha/D-4$ confined to the limit $D \rightarrow 4$ \cite{Glav:2020a}.
Howbeit, Metin et al. passed contradictory remarks against this method in \cite{Mg:2020a} by revealing the lack of continuity
of the EGB theory at $D=4$. Shortly after that, similar comments regarding the rescaled term $1/D-4$ were addressed by  Arrechea et al. in the letter \cite{Jul:2020a} and observed some ill-defined terms affecting the metric perturbations.
Although, the authors have now constructed positive arguments in response
to the linear perturbations in the 4D-EGB background \cite{Jul:2020b}, hence, we shall employ linear weak-field approximations to
study the lensing phenomenon in 4D-EGB gravity.

This freshly proposed theory of 4D-EGB has up till now been studied for various astrophysical observations to uncover the intricate features of the space-time gravity. By considering the indicated gravity the gravitational collapse of a homogenous dust sphere is tactfully presented in \cite{Daniel:2020a}.  The
superradiance and stability along with the quasinormal modes of a charged 4D-EGB black hole are worked out in \cite{Zhang:2020a} and the EGB model endowed
with a scalar field \cite{Gilb:2019a} is used to explore its stability by utilizing the odd parity perturbations method.
In Refs \cite{Yupeng:2020a,Ming:2020a} the authors investigated the dynamics of spinning test particles, the innermost stable
circular orbits and shadow cast in surroundings of a static 4D-EGB gravity; whereas a cognitive analysis for a rotating
counterpart has been reviewed in \cite{Rahul:2020a}. The validity of strong cosmic censorship for charged de Sitter black hole is studied in the context
of scalar and electromagnetic perturbation \cite{Akash:2020a}. The quasinormal modes of the Dirac field and the perturbative as well as the non-perturbative
 modes are, respectively, inspected in the \cite{Churi:2020a,Almen:2020a}. Various aspects of black holes such as gravitational lensing,
 optical properties, thermodynamics and particle acceleration are profoundly examined in \cite{Islam:2020a,Abu:2015d,Hedge:2020a,
 Mansr:2020a,Kumara:2020a}. One may get relevant information concerning the 4D-EGB gravity from the research performed in
 the Refs \cite{Li:2020a,Wei:2020b,Dhrm:2020a,Wei:2020c,Moha:2020a,Met:2020a,Robie:2020a,Juli:2020a,Tian:2020a,Jame:2020a,
Yipng:2020a,Tsut:2020a,Fern:2020a,Abu:2020f,Shay:2020a,Maha:2020a,Feng:2021a,Olivera:2009a}.

The paper is organised as follows. In Sec.~(\ref{metric}) the deflection angle of the massless particles is precisely studied
in the light of a weak-field approximation by considering the EGB gravity embedded in three different
mediums, i.e, uniform plasma, singular isothermal sphere and a non-singular isothermal sphere.
Sec.~(\ref{Magnification}) provides a detailed logical reasoning about the apparent magnification of the source star.
Finally, in Sec.~(\ref{con}) we have summarized our main results.

\section{Weak-field lensing in the presence of plasma}\label{metric}
The action of D-dimensional Einstein-Gauss-Bonnet theory with a redefined coupling constant $\alpha\rightarrow\frac{\alpha}{D-4}$ is expressed
by the relation \cite{Glav:2020a,Zhang:2020a},

\begin{eqnarray}
\mathcal{S}&=\frac{1}{16\pi}\bigintssss d^Dx\sqrt{-g}\Big(R+\frac{\alpha}{D-4}\mathcal{G}\Big),
\end{eqnarray}
where $\alpha$ is a dimensionless Gauss–Bonnet (GB) coupling parameter and $\mathcal{G}$ is the
Gauss–Bonnet invariant defined by the expression
\begin{align}
\mathcal{G}=R^{\mu\nu\eta\rho}R_{\mu\nu\eta\rho}-4R^{\mu\nu}R_{\mu\nu}+R^2,
\end{align}
$R$ is the Ricci scalar, $R_{\mu\nu}$ and $R_{\mu\nu\eta\rho}$ denote, respectively, the Ricci and
Riemann tensors.
The action $\mathcal{S}$ in a 4 dimensional analysis yields the line-element of a non rotating 4D-EGB gravity
in the form

\begin{align}
 d s^2&=-f(r)dt^2+f^{-1}(r)dr^2+r^2(d\theta^2+\sin^2\theta d\phi^2),
 \end{align}
 where
 \begin{align}
 f(r)=1+\frac{r^2}{2\alpha}\bigg(1-\sqrt{1+\frac{4\alpha R_s}{r^3}}\bigg),
 \end{align}

here $R_s=2M$, the value of GB parameter $\alpha/M^2$ lies in the range [-8,1] \cite{Abu:2020f}. Note that, $\alpha>1$ corresponds to naked singularities
and $\alpha<-8$ leads to complex-valued metric in the outer region of the event horizon\cite{Ming:2020a}. Moreover,
Schwarzschild metric is recovered when $\alpha\rightarrow0$.
We shall take the series expansion of $f(r)$ upto order $O(R^3_s)$ for a more exhaustive evaluation,
\begin{align}
f(r)&=1-\frac{R_s}{r}+\frac{\alpha R^2_s}{r^4}.
\end{align}

In this section our main concern is to unravel the effects of gravitational lensing
in the background of 4D-EGB gravity surrounded by a plasma considering a weak-field approximation defined as follows,

\begin{align}
g_{\alpha\beta}=\eta_{\alpha\beta}+h_{\alpha\beta},
\end{align}
where $\eta_{\alpha\beta}$ and $h_{\alpha\beta}$ connote the Minkowski metric and perturbation metric, respectively.
\begin{eqnarray}
&&\eta_{\alpha\beta} = {\rm diag} (-1, 1, 1, 1)\ ,\nonumber\\
&& h_{\alpha\beta} \ll 1, \quad  h_{\alpha\beta} \rightarrow 0 \quad {\rm under } \quad x^{\alpha}\rightarrow \infty\ , \nonumber \\
&& g^{\alpha\beta}=\eta^{\alpha\beta}-h^{\alpha\beta},\ \ \ h^{\alpha\beta}=h_{\alpha\beta}\ .
\end{eqnarray}
Before proceeding any further we shall briefly recall the layout presented in \cite{Bin:2010a,Abu:2013a,Abu:2017a}
to elicit the general expression of the angle of deflection. The correlation between the phase velocity
$\textit{\textbf{v}}$ and the 4-momentum $\textit{\textbf{p}}^\alpha$ considering a static case is expressed as \cite{Synge:1960b},

\begin{align}\label{phasevp}
\frac{c^2}{\textit{\textbf{v}}^2}&=n^2=1+\frac{\textit{\textbf{p}}_\alpha \textit{\textbf{p}}^\alpha}{(\textit{\textbf{p}}^0\sqrt{-g_{00}})^2}.
\end{align}
By taking into account a dispersive medium, Synge \cite{Synge:1960b} remodeled
the Fermat’s least action principle to describe the photon trajectories. Henceforth, the variational principle $\partial(\int \textit{\textbf{p}}_\alpha dx^\alpha)=0$ was applied with the condition,

\begin{align}\label{Hamilton}
H(x^\alpha,\textit{\textbf{p}}_\alpha)=\frac{1}{2}[g^{\alpha\beta}\textit{\textbf{p}}_\alpha \textit{\textbf{p}}_\beta-(n^2-1)(\textit{\textbf{p}}_0
\sqrt{-g_{00}})^2].
\end{align}
The above expression governs the equations of motion by the following system of differential equations
$\frac{dx^\alpha}{d\lambda}=\frac{\partial H}{\partial \textit{\textbf{p}}_\alpha},
\frac{d\textit{\textbf{p}}_\alpha}{d\lambda}=-\frac{\partial H}{\partial x^\alpha}$, here $\lambda$ is the affine parameter.
 It is of utmost importance to define the refractive index $n$ properly in order to study the plasma effects
clearly in the black hole  vicinity, therefore by the implication of \cite{Bin:2010a} we define it as

\begin{align}\label{refractiveindex}
n^2=1-\frac{{\omega_{e}}^2}{[\omega(x^i)]^2}, \quad  \omega_{e}^2&=\frac{4\pi e^2 N(x^i)}{m}=K_{e}N(x^i).
\end{align}
Here, $\omega_e$ is the plasma electron frequency and $\omega(x^i)$ is a space coordinate function termed as the photon frequency.
The notations $e$, $m$ and $N(x^i)$ denote the charge, mass and electron concentration, correspondingly. The validity of the inequality, $\omega^2>{\omega_{e}}^2$ is central to the propagation of light through the plasma medium.
While considering a non-rotating gravitational field with a static medium, the photon energy reads  \cite{Synge:1960b}

\begin{align}\label{photonenergy}
\textit{\textbf{p}}^0\sqrt{-g_{00}}=-\frac{1}{c}\hbar \omega(x^i).
\end{align}
Using (\ref{photonenergy},\ref{refractiveindex}) the scalar $H(x^\alpha,\textit{\textbf{p}}_\alpha)$ takes the form

\begin{align}\label{Hamiltoneq3}
H(x^\alpha,\textit{\textbf{p}}_\alpha)=\frac{1}{2}\big[g^{\alpha\beta}\textit{\textbf{p}}_\alpha \textit{\textbf{p}}_\beta+\frac{\omega^2_e \hbar^2}{c^2}\big],
\end{align}
where $\hbar$ is the Planck’s constant. Generally, for any arbitrary medium, photons in a flat space-time move along a straight path, while on the
other hand, bent trajectories are followed in a curved space-time. Thus, we assume the motion specifically
along z-axis and take in the null approximations \cite{Bin:2010a} to avoid any small deviations from the straight path.
In this case the components of the 4-momentum are as below

\begin{align}\label{momentumcomp}
\textit{\textbf{p}}^\alpha=\bigg(\frac{\hbar \omega}{c},0,0,\frac{n \hbar \omega}{c}\bigg), \quad \textit{\textbf{p}}_\alpha=\bigg(-\frac{\hbar \omega}{c},0,0,
\frac{n \hbar \omega}{c}\bigg).
\end{align}

Note that, in the forthcoming discussion we shall utilize special notations at infinity;
$\omega(\infty)=\omega$, $\omega_e(\infty)=\omega_0$  and $n(\infty)=\sqrt{1-\frac{\omega^2_0}{\omega^2}}=n$ .
Since we are considering a diagonal metric therefore the components of the metric tensor
$g_{\alpha\beta}$ dissolves for all $\alpha\neq \beta$. Hence, after using (\ref{Hamiltoneq3})
we get the following set of equations

\begin{align}\label{system}
\frac{dx^i}{d\lambda}&=g^{ij}\textit{\textbf{p}}_j, \nonumber\\
\frac{d\textit{\textbf{p}}_i}{d\lambda}&=-\frac{1}{2}g^{lm}_{,i}\textit{\textbf{p}}_l\textit{\textbf{p}}_m-\frac{1}{2}g^{00}_{,i}\textit{\textbf{p}}^2_0-\frac{1}{2}\frac{\hbar^2}
{c^2}K_eN,i.
\end{align}
After applying the null approximation (\ref{system}) reduces to

\begin{align}
\frac{dz}{d\lambda}=\frac{n\hbar \omega}{c}.
\end{align}
The photon momentum in terms of 3-dimensional standard unit vector $\textit{\textbf{u}}^i$ =$\textit{\textbf{u}}_i$ = (0, 0, 1)
can be expressed as

\begin{align}\label{momentcomp}
\textit{\textbf{p}}_i=\frac{n\hbar \omega}{c}(0,0,1)=\frac{n\hbar \omega}{c}\textit{\textbf{u}}_i.
\end{align}
By substituting the above equation in (\ref{system}) we get,

\begin{align}\label{Motion}
\frac{d}{d\lambda}\bigg(\frac{n\hbar \omega}{c}\textit{\textbf{u}}_i\bigg)&=-\frac{1}{2}g^{lm}_{,i}\textit{\textbf{p}}_l\textit{\textbf{p}}_m-
\frac{1}{2}g^{00}_{,i}\textit{\textbf{p}}^2_0-\frac{1}{2}\frac{\hbar^2}{c^2}K_eN,i.
\end{align}
Owing to the preceding assumption made i.e, the motion takes place
only along z axis, we are confined to consider only those components of the unit vector
which are perpendicular to the initial direction of propagation (see \cite{Bin:2010a}). Finally, complying with the
the null approximation in addition to a weak gravitational field (\ref{Motion}) appears as

\begin{align}\label{weakfield}
\frac{d\textit{\textbf{u}}_i}{dz}=\frac{1}{2}\bigg(h_{33,i}+\frac{1}{n^2}h_{00,i}-\frac{1}{n^2\omega^2}K_eN_{,i}\bigg), \quad i=1,2.
\end{align}
The deflection angle is basically defined by $\hat{a}=\textit{\textbf{u}}_{+\infty}-\textit{\textbf{u}}_{-\infty}$, thus, (\ref{weakfield}) leads us to
a general expression for it as below

\begin{eqnarray}\label{alphak}
&&\hat{\alpha}_i= \frac{1}{2} \int_{-\infty}^\infty \bigg(h_{33_,i} +
\frac{ \omega^2}{\omega^2-\omega_e^2}h_{00_,i}-\frac{K_e }{\omega^2-\omega_e^2}N{_,i}\bigg) dz \ .\nonumber \\
\end{eqnarray}
The $\mp$ signs of $\hat{\alpha}_i$ determines the deflection towards and away from the central object, respectively.
At large $r$ we have $\frac{R_s}{r}>>(\frac{R_s}{r})^2$, thereby the black hole metric is approximated to

\begin{align}
ds^2=ds^2_0+\bigg(\frac{R_s}{r}-\frac{\alpha R^2_s}{r^4}\bigg)dt^2+\bigg(\frac{R_s}{r}-\frac{\alpha R^2_s}{r^4}\bigg)dr^2
\end{align}
where $ds^2_0=-dt^2+dr^2+r^2(d\theta^2+\sin^2\theta d\phi^2)$. In the Cartesian coordinates the components $h_{\alpha\beta}$ can be written as
\begin{eqnarray}
 h_{00}&=&\bigg(\frac{R_s}{r}-\frac{\alpha R^2_s}{r^4}\bigg), \nonumber \\  h_{ik}&=&\bigg(\frac{R_s}{r}-\frac{\alpha R^2_s}{r^4}\bigg)n_{i}n_{k}\ ,\nonumber \\
 h_{33}&=&\bigg(\frac{R_s}{r}-\frac{\alpha R^2_s}{r^4}\bigg)\cos^2x \ ,
\end{eqnarray}\label{h}
where $\cos x=z/\sqrt{b^2+z^2}$ and $r=\sqrt{b^2+z^2}$, $b$ is the impact parameter signifying the closest approach of the photons to the black hole.
Using the above mentioned expressions in the formula (\ref{alphak}) one can compute the light deflection angle with respect to $b$
for a black hole surrounded by plasma

\begin{figure}[h!]
\includegraphics[scale=0.17]{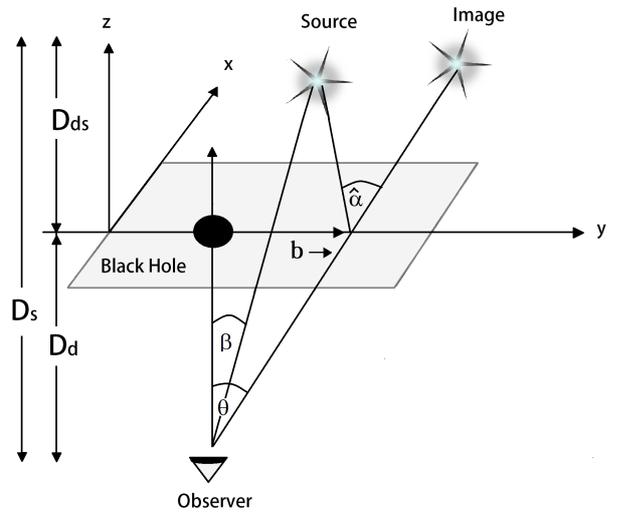}
\caption{Schematic representation of the gravitational lensing system typically based on the black hole lens, source star
and the image formed.}\label{Lensingsystem}
\end{figure}

\begin{align}
\hat{\alpha}_{b}&=\bigintssss_{-\infty}^{\infty}\frac{b}{2r}
\Bigg(\partial_r \bigg(\bigg(\frac{R_s}{r}-\frac{\alpha R^2_s}{r^4}\bigg)\cos^2x\bigg)+
\partial_r\bigg(\frac{R_s}{r}-\frac{\alpha R^2_s}{r^4}\bigg)\frac{\omega^2}{\omega^2-\omega^2_e}\nonumber\\&
-\frac{K_e}{\omega^2-\omega^2_e}\partial_r N\Bigg)dz.
\label{alfa}
\end{align}

In the light of foregoing discussion,
we can easily examine the impact of different plasma mediums on the photon deflection angle as follows:

\begin{figure}[h!]
 \begin{center}
   \includegraphics[scale=0.38]{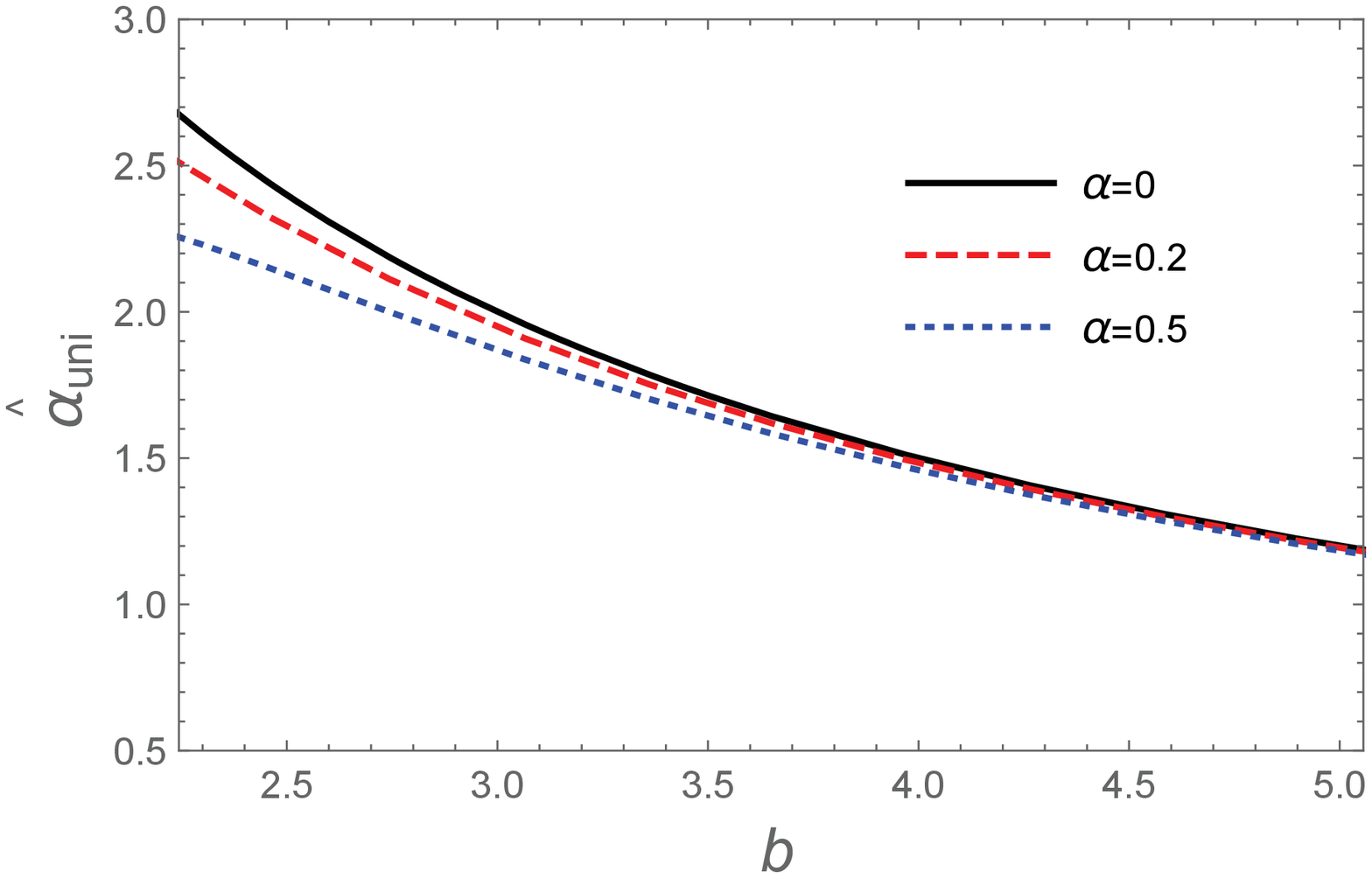}
   \includegraphics[scale=0.38]{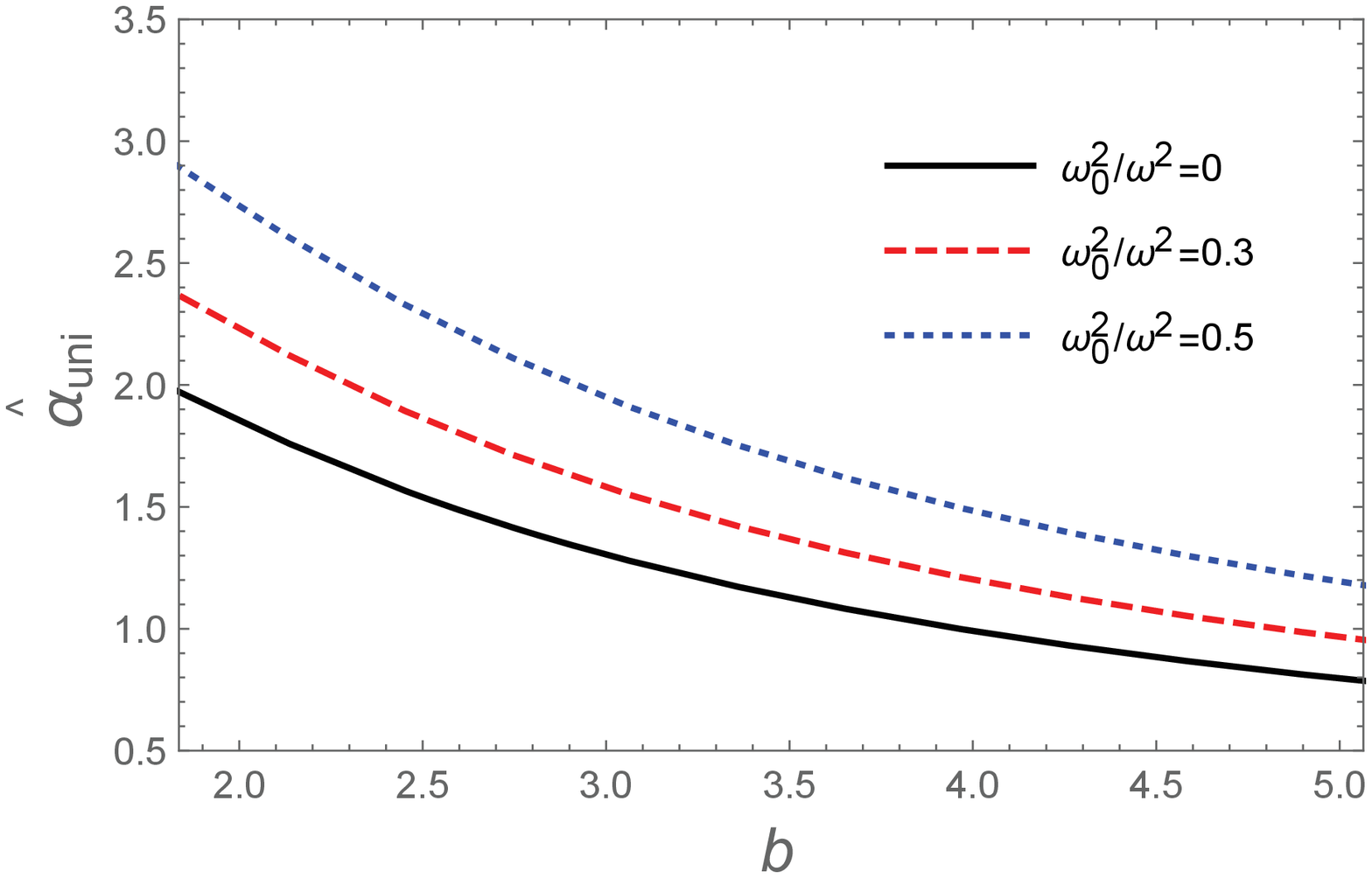}
  \end{center}
\caption{Plot of the deflection angle $\hat{\alpha}_{\mathrm{uni}}$ as a function of the impact
parameter $b$ for $\frac{\omega^2_{0}}{\omega^2}=0.5$ (upper panel)and $\alpha=0.5$ (lower panel).}\label{angleuni1}
\end{figure}

\begin{figure}[h!]
\centering
  \includegraphics[align=t,scale=0.38]{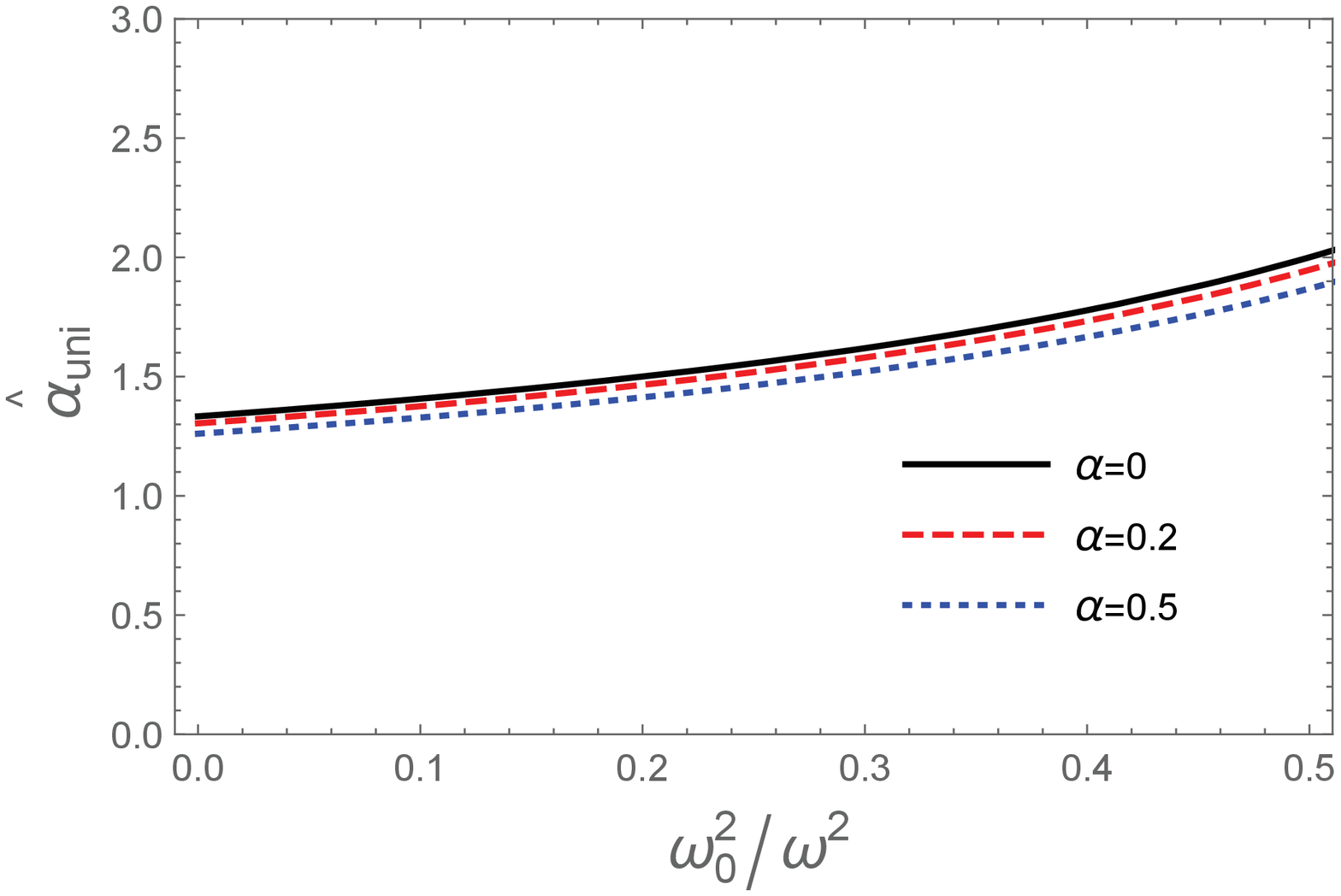}
  \includegraphics[align=t,scale=0.38]{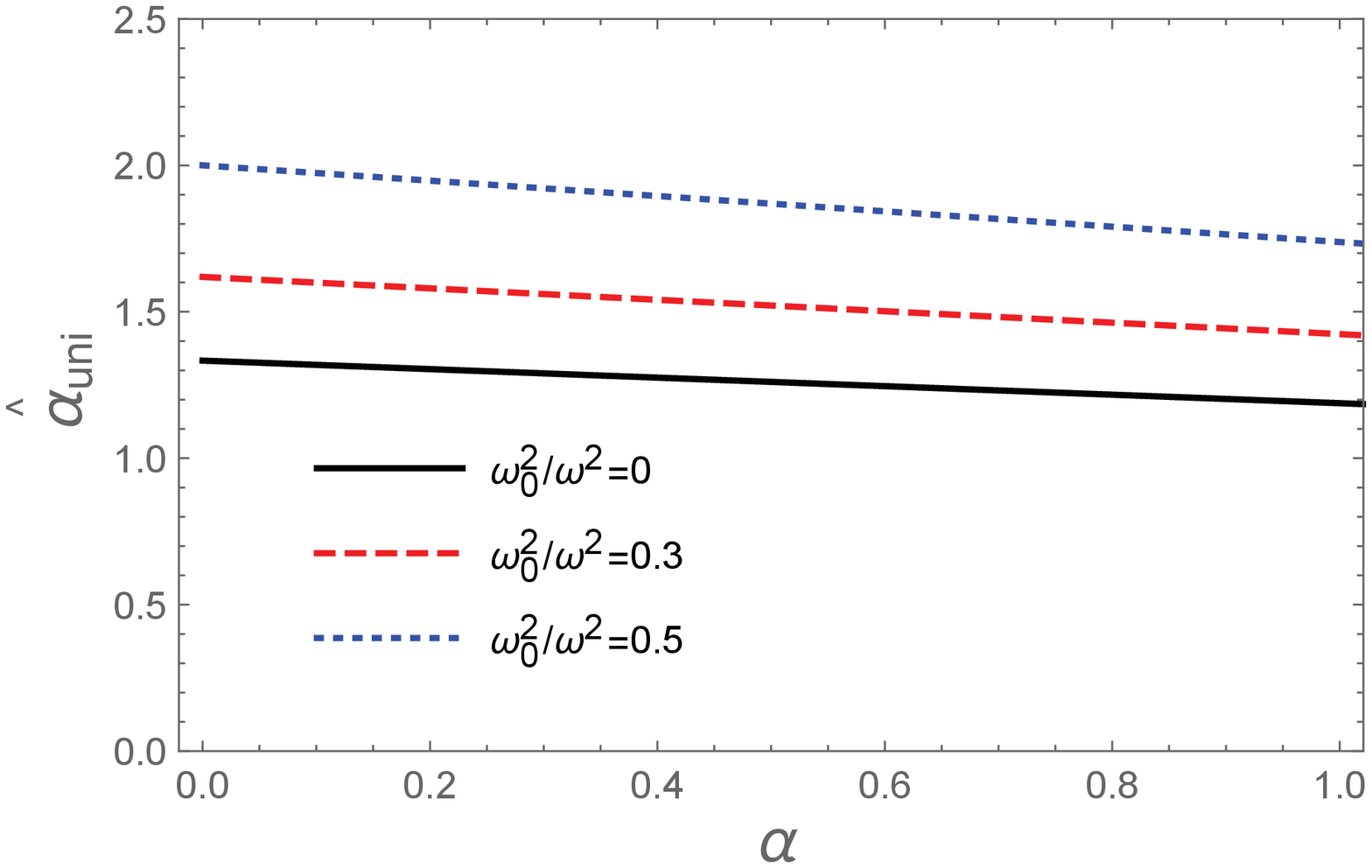}
\caption{Plot of the deflection angle $\hat{\alpha}_{\mathrm{uni}}$ as a function of
$\frac{\omega^2_{0}}{\omega^2}$ (upper panel) and $\alpha$ (lower panel) with a fixed impact parameter $b=3$.}\label{angleuni2}
\end{figure}
\subsection{$\mathbf{Uniform}$ $\mathbf{Plasma}$}
First of all, we consider the photon geodesics when plasma is uniformly distributed
in the black hole surroundings. In homogenous plasma medium the refractive index $n$ particulary counts on the
photon frequency with $\omega_0$ as a constant quantity which ultimately leads to the approximation $1-n<<\frac{\omega_0}{\omega}$ \cite{Car:2018a}.
Subsequently, the corresponding constraints annihilate the term $\partial_r N$ and the angle of deflection takes the form

\begin{align}\label{unialpha}
\hat{\alpha}_{\mathrm{uni}}&=\bigg(\frac{R_s}{b}-\frac{3\pi \alpha R^2_s}{16b^4}\bigg)+\bigg(\frac{R_s}{b}-\frac{3\pi
 \alpha R^2_s}{4b^4}\bigg)\frac{1}{\big(1-\frac{\omega^2_0}{\omega^2}\big)}.
\end{align}

Fig. (\ref{angleuni1}) illustrates the plots of the photon deflection angle $\hat{\alpha}_{\mathrm{uni}}$ as a function of the
impact parameter $b$ for various coupling constant $\alpha$ (upper panel) and plasma parameters $\frac{\omega^2_{0}}{\omega^2}$ (lower panel).
An increase is examined in the deflection angle for smaller values of the impact parameter $b$, which means that
a massless particle passing too close to the black hole surroundings basically enhances its deviating tendency.
Fig. (\ref{angleuni2}) is a visualization of the deflection angle distinctively with respect to $\frac{\omega^2_{0}}{\omega^2}$ and $\alpha$.
The deflection angle is maximum due to high plasma distribution (upper panel) and is seen to be strictly decreasing
against an increasing coupling parameter $\alpha$ (lower panel), for instance, taking $\alpha=0$, the Schwarzschild
gravity ensures the highest degree of deviation $\hat{\alpha}_\mathrm{uni}$. We deduce that, as expected, the existence of plasma in the black hole vicinity, contrariwise to the vacuum case $\frac{\omega^2_{0}}{\omega^2}=0$, contributes to
the photon motion.

\subsection{$\mathbf{Singular}$ $\mathbf{Isothermal}$ $\mathbf{sphere}$}
A singular isothermal sphere (SIS) is the most favourable model to comprehend the peculiar features of gravitational lensed photons. It
was primarily introduced in \cite{Chnd:1939a,Jbin:1987a} to explore the len's property of the galaxies and clusters.
Generally, SIS is a spherical gas cloud with a singularity located at its center where the density tends to
infinity. The density distribution of a SIS is given by
\begin{align}
\rho(r)=\frac{\sigma^2_{v}}{2\pi r^2},\label{rho}
\end{align}
where $\sigma^2_{v}$ refers to a one-dimensional velocity dispersion. The plasma concentration
admits the following analytic expression

\begin{align}
N(r)=\frac{\rho(r)}{\kappa m_p},\label{conelec}
\end{align}
here $m_p$ is the proton mass and $\kappa$ is a dimensionless
constant coefficient generally associated to the dark matter universe \cite{Bin:2010a}.
Utilizing (\ref{refractiveindex},\ref{rho},\ref{conelec}) the plasma frequency takes the form

\begin{align}
\omega^2_{e}=K_eN(r)=\frac{K_e\sigma^2_{v}}{2\pi \kappa m_p r^2}.
\end{align}
We reckon with the above mentioned  properties of the SIS and compute the angle of deflection $\hat{\alpha}_{SIS}$ as below \cite{Bin:2010a},

\begin{align}
\hat{\alpha}_\mathrm{SIS}=\Bigg(\frac{2R_s}{b}-\frac{15\pi \alpha R^2_s}{16b^4}\Bigg)+\frac{R^2_s\omega^2_c}{b^2\omega^2}
\Bigg(\frac{1}{2}-\frac{2R_s}{3b\pi}+\frac{5\alpha R^2_s}{8b^4}\Bigg). \label{alphasis}
\end{align}
These calculations brings up a supplementary plasma constant $\omega^2_c$ which has the following analytic expression.

\begin{align}
\omega^2_c=\frac{\sigma^2_{v} K_e}{2\kappa m_p R^2_s}.
\end{align}

\begin{figure}[h!]
 \begin{center}
   \includegraphics[scale=0.38]{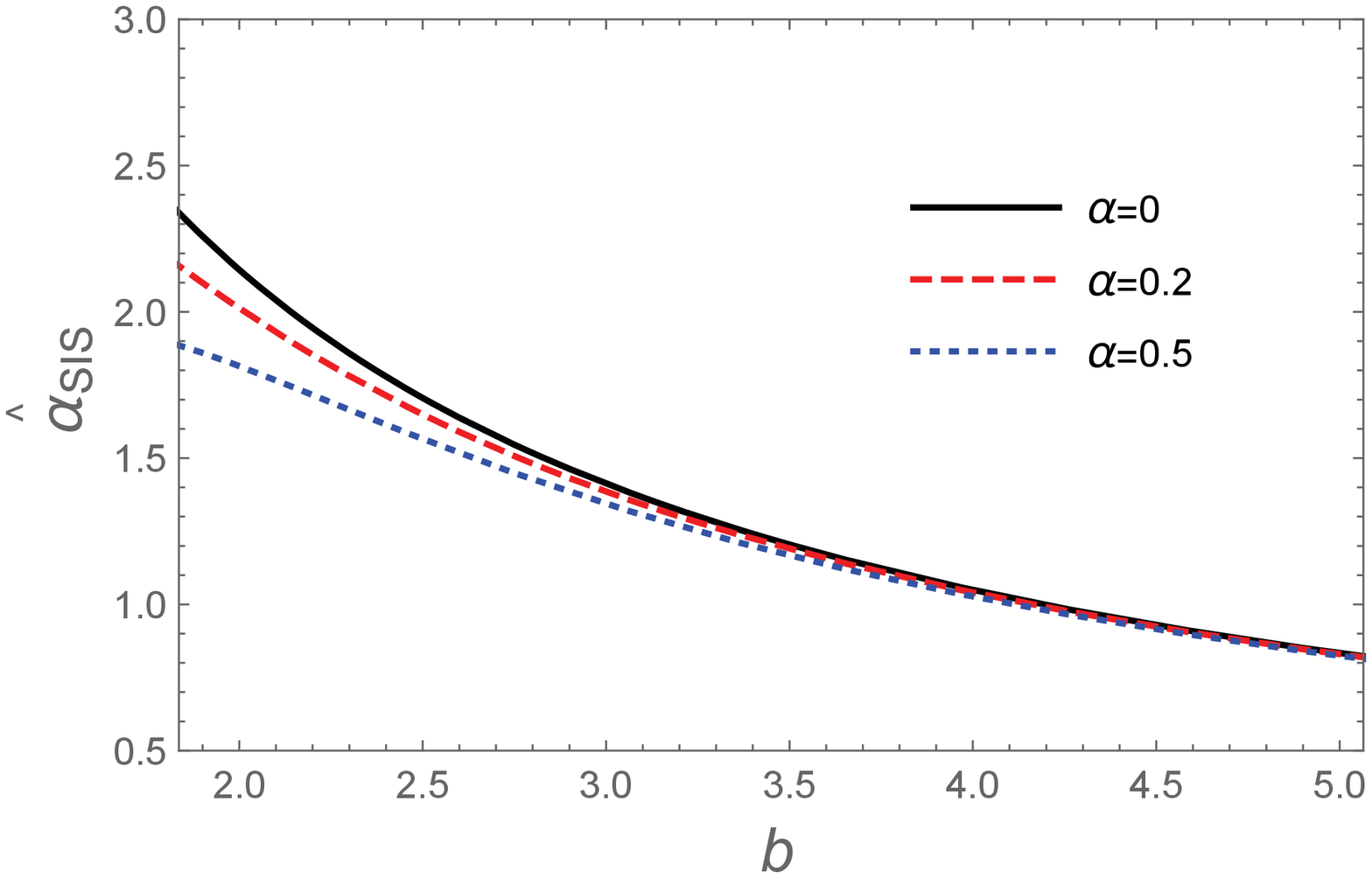}
   \includegraphics[scale=0.38]{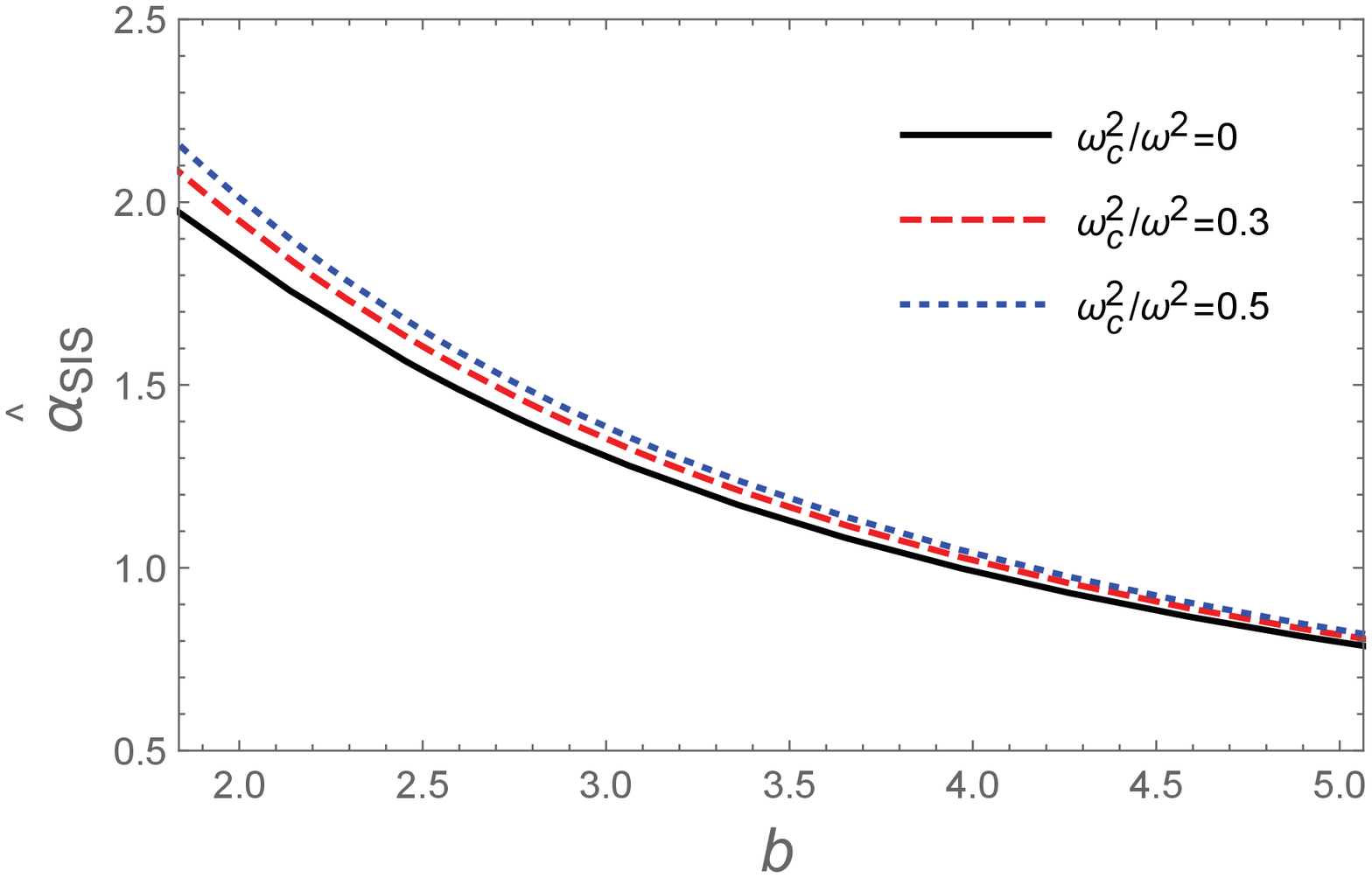}
  \end{center}
\caption{Plot of the deflection angle $\hat{\alpha}_\mathrm{SIS}$ as a function of the impact
parameter $b$ for $\frac{\omega^2_{0}}{\omega^2}=0.5$ (upper panel) and $\alpha=0.5$ (lower panel).}\label{anglesis1}
\end{figure}

\begin{figure}[h!]
 \begin{center}
   \includegraphics[align=t,scale=0.38]{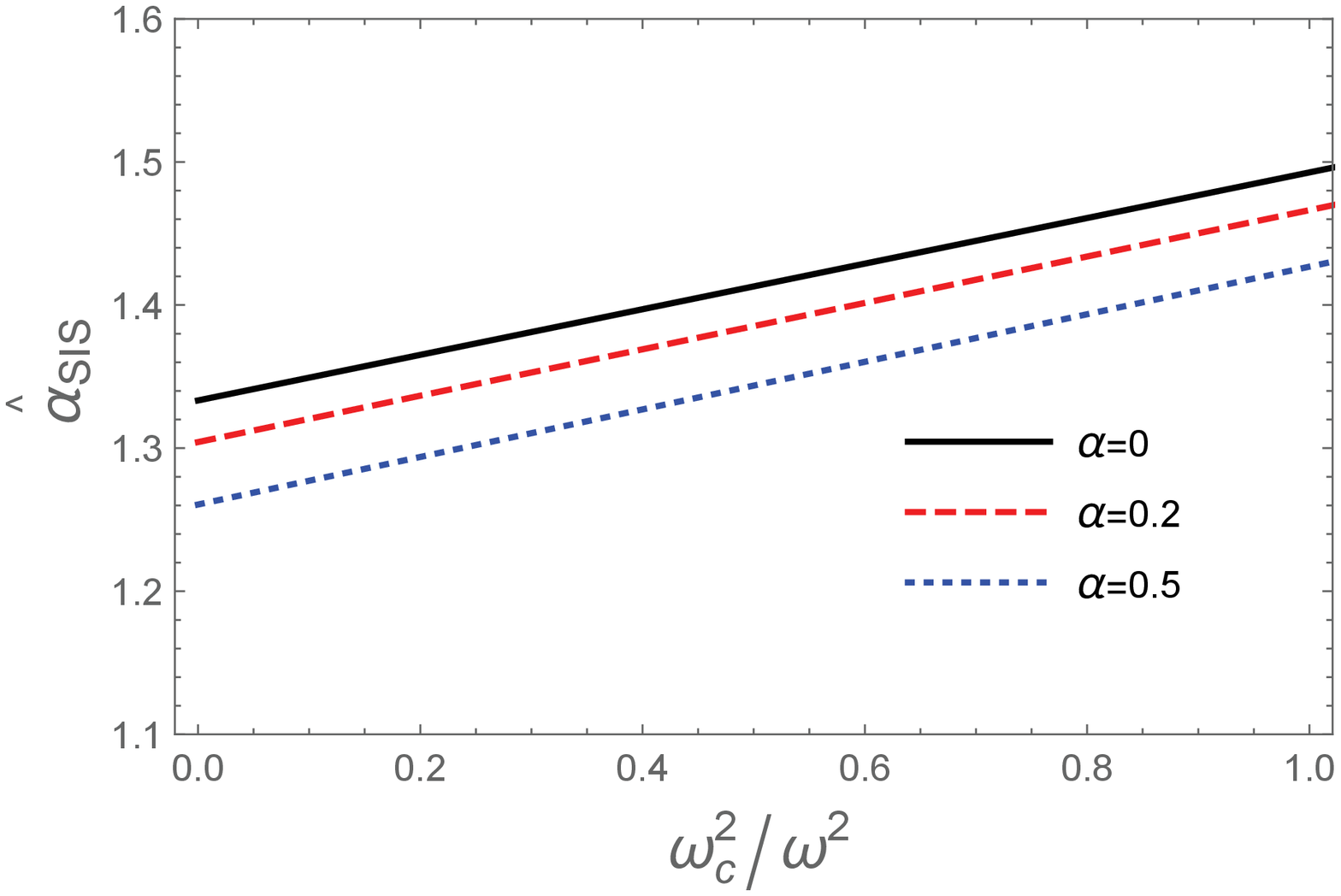}
   \includegraphics[align=t,scale=0.38]{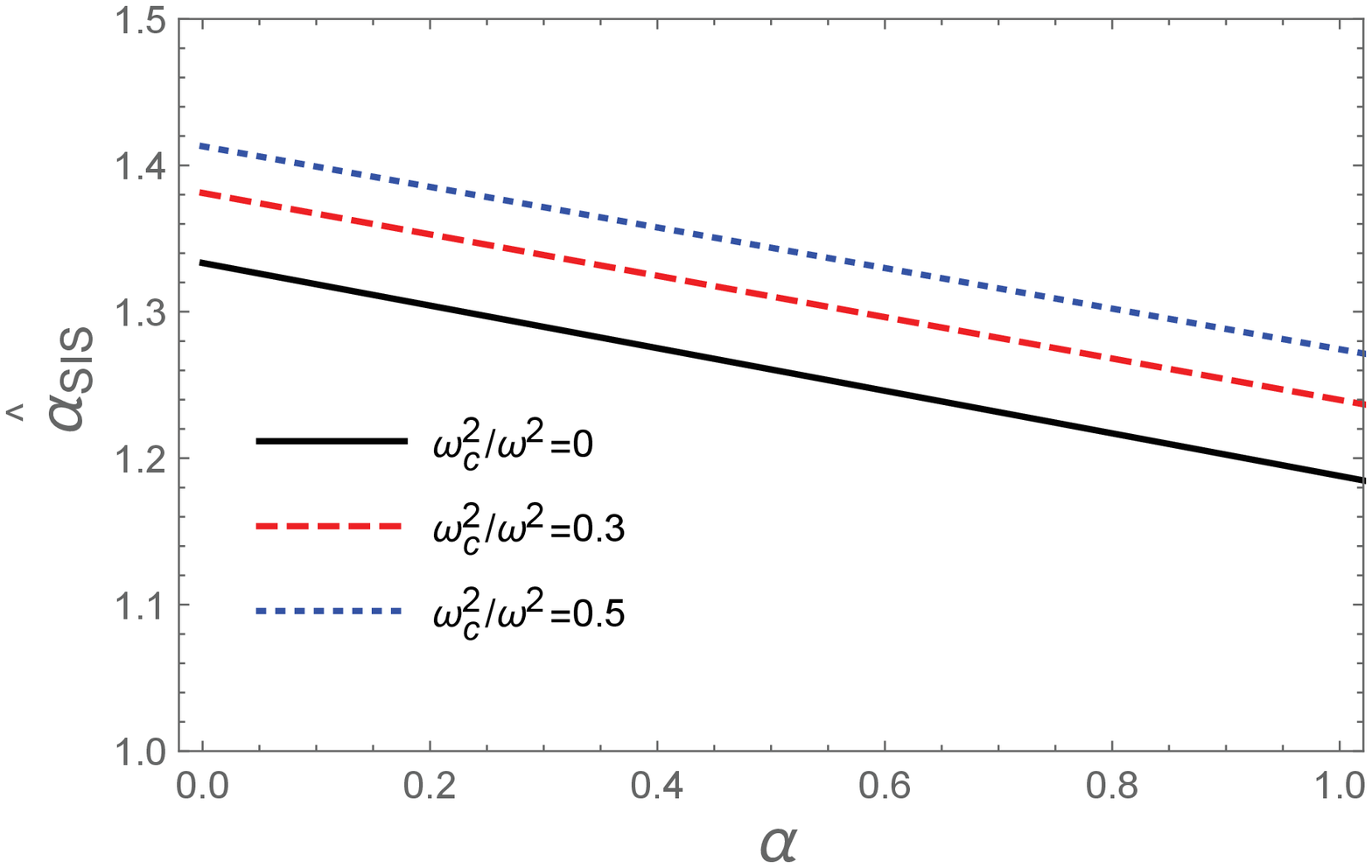}
  \end{center}
\caption{Plot of the deflection angle $\hat{\alpha}_\mathrm{SIS}$ as a function of $\frac{\omega^2_{c}}{\omega^2}$ (upper panel) and $\alpha$ (lower panel)
for a fixed impact parameter $b=3$.}\label{anglesis2}
\end{figure}

In order to assimilate the influence of SIS on the photon trajectory we plotted
the deflection angle $\hat{\alpha}_{\mathrm{SIS}}$ as a function of the impact parameter $b$, see Fig. (\ref{anglesis1}),
interestingly, we see that the uniform plasma and SIS medium share common features regarding the parameter $b$.
Note that, the quantity $\frac{\omega^2_{c}}{\omega^2}$ identifies the distribution of SIS in the black hole vicinity,
thus we detect the photon sensitivity to the specified parameter along with the coupling constant parameter $\alpha$, by means of a graphical analysis in Fig. (\ref{anglesis2}). We examined that $\hat{\alpha}_{\mathrm{SIS}}$ increases when $\frac{\omega^2_{c}}{\omega^2}$ increases (upper panel) and, conversely, $\hat{\alpha}_{\mathrm{SIS}}$ decreases when $\alpha$ increases (lower panel). Hence,
the presence of SIS in the black hole surroundings to some extent affects the intervening massless particles.

\subsection{$\mathbf{Non}$-$\mathbf{Singular}$ $\mathbf{Isothermal}$ $\mathbf{gas}$ $\mathbf{sphere}$}
Now we further proceed to study the motion of photons considering a non-singular isothermal sphere (NSIS) which is a more reasonable and physical
setup for the analysis. Unlike the SIS, in this lens model the singularity is bounded by a finite core at the origin of the gas cloud whereby
the density distribution is defined as \cite{Hind:1987a,Wu:1996a}
\begin{align}
\rho(r)=\frac{\sigma^2_{v}}{2\pi (r^2+r^2_c)}=\frac{\rho_0}{(1+\frac{r^2}{r^2_c})}, \quad \rho_0=\frac{\sigma^2_{v}}{2\pi r^2_c}, \label{rhonsis}
\end{align}
here the core radius is represented by $r_c$. The plasma concentration for NSIS using (\ref{conelec}) becomes

\begin{align}
N(r)=\frac{\sigma^2_{v}}{2\pi \kappa m_p (r^2+r^2_c)}. \label{Nrnsis}
\end{align}
We compute the plasma frequency from (\ref{refractiveindex},\ref{rhonsis},\ref{Nrnsis}) as follows
\begin{align}
\omega^2_e=\frac{K_e\sigma^2_v}{2\pi\kappa m_p(r^2+r^2_c)}.
\end{align}
The angle of deflection obtained by the deviation of photons in NSIS gravitational lens setup in accordance with the properties in the latter discussion
is as below
\begin{align}
\hat{\alpha}_\mathrm{NSIS}&=\Bigg(\frac{2R_s}{b}-\frac{15\pi \alpha R^2_s}{16b^4}\Bigg)+\frac{R^2_s\omega^2_c}{\omega^2}\Bigg(\frac{R_s}{b\pi r^2_c}+
\frac{b}{2(\sqrt{b^2+r^2_c})^3}\nonumber\\&-\frac{bR_s\tanh^{-1}\frac{r_c}{\sqrt{b^2+r^2_c}}}{\pi r^3_c\sqrt{b^2+r^2_c}}
-\frac{\alpha R^2_s}{r^2_c}\Bigg(\frac{2}{r^4_c}+\frac{3}{4b^4}-\frac{1}{b^2r^2_c}\nonumber\\&-\frac{2b}{r^4_c\sqrt{b^2+r^2_c}}\Bigg)\Bigg).
\end{align}

\begin{figure}[h!]
 \begin{center}
   \includegraphics[scale=0.38]{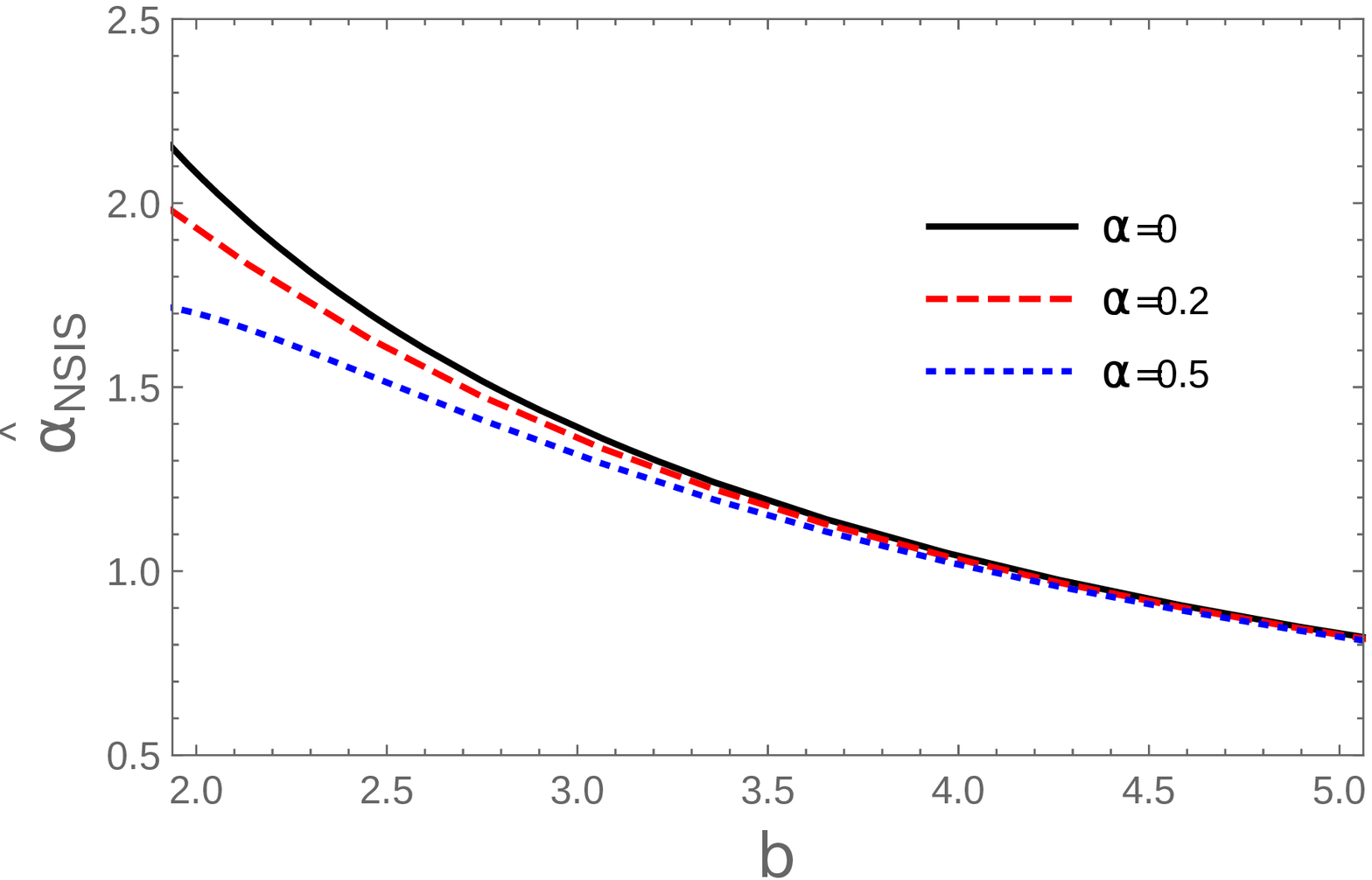}
   \includegraphics[scale=0.38]{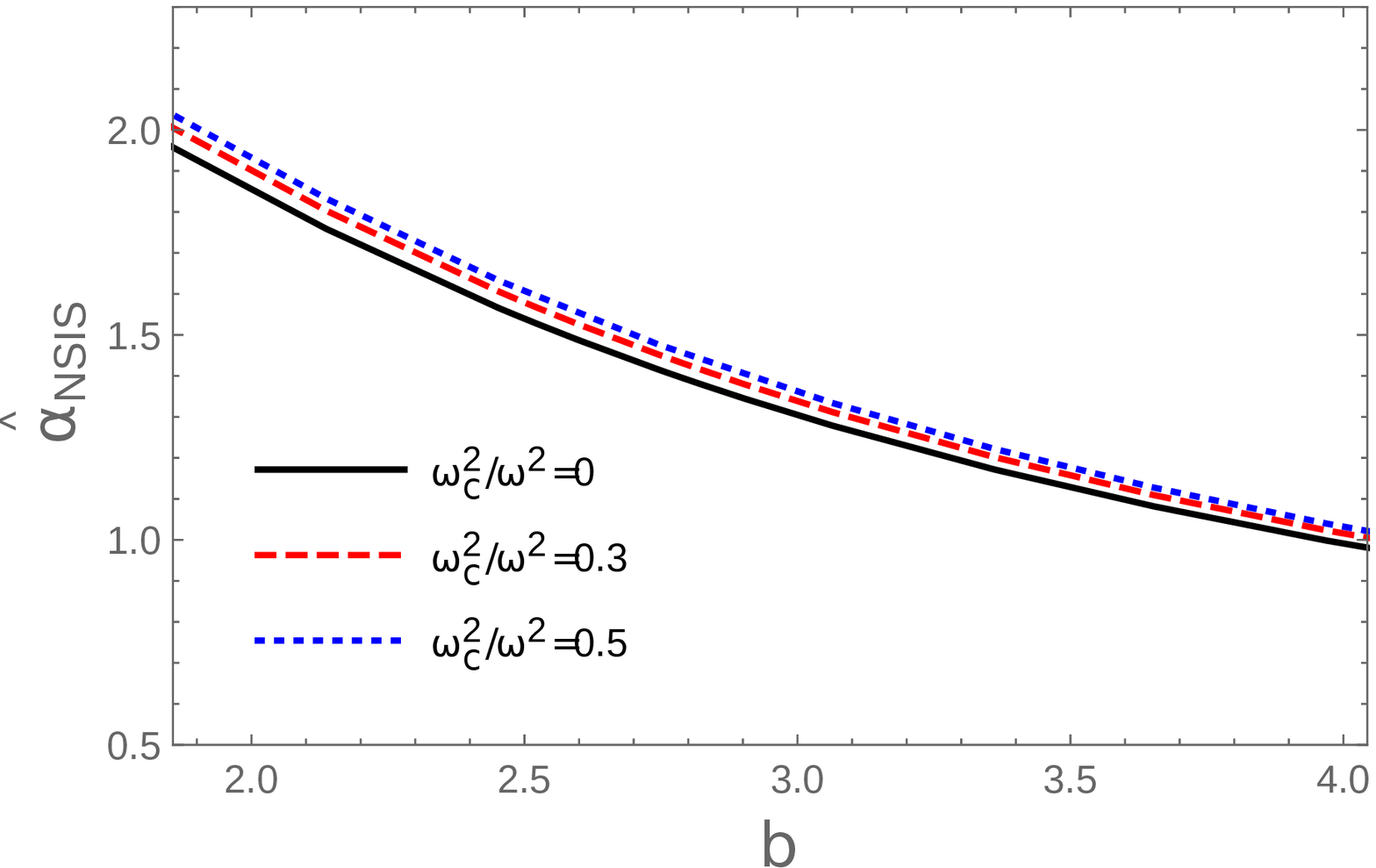}
  \end{center}
\caption{Plot of the deflection angle $\hat{\alpha}_\mathrm{NSIS}$ as a function of the impact
parameter $b$ for $\frac{\omega^2_{c}}{\omega^2}=0.5$
(upper panel) and $\alpha=0.5$ (lower panel) with a fixed parameter $r_c=3$ .}\label{angleNsis1}
\end{figure}

\begin{figure}[h!]
 \begin{center}
   \includegraphics[align=t,scale=0.38]{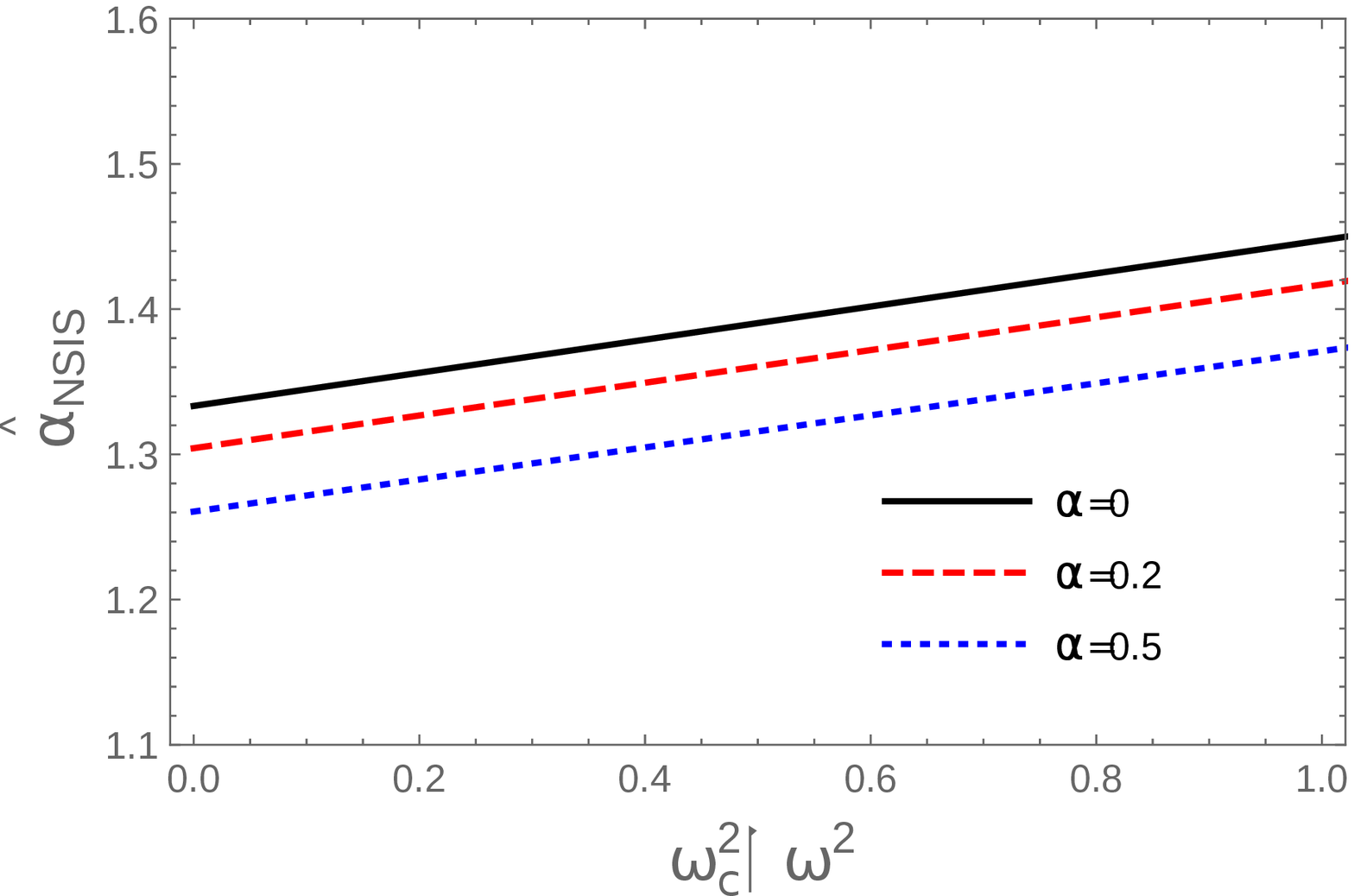}
   \includegraphics[align=t,scale=0.38]{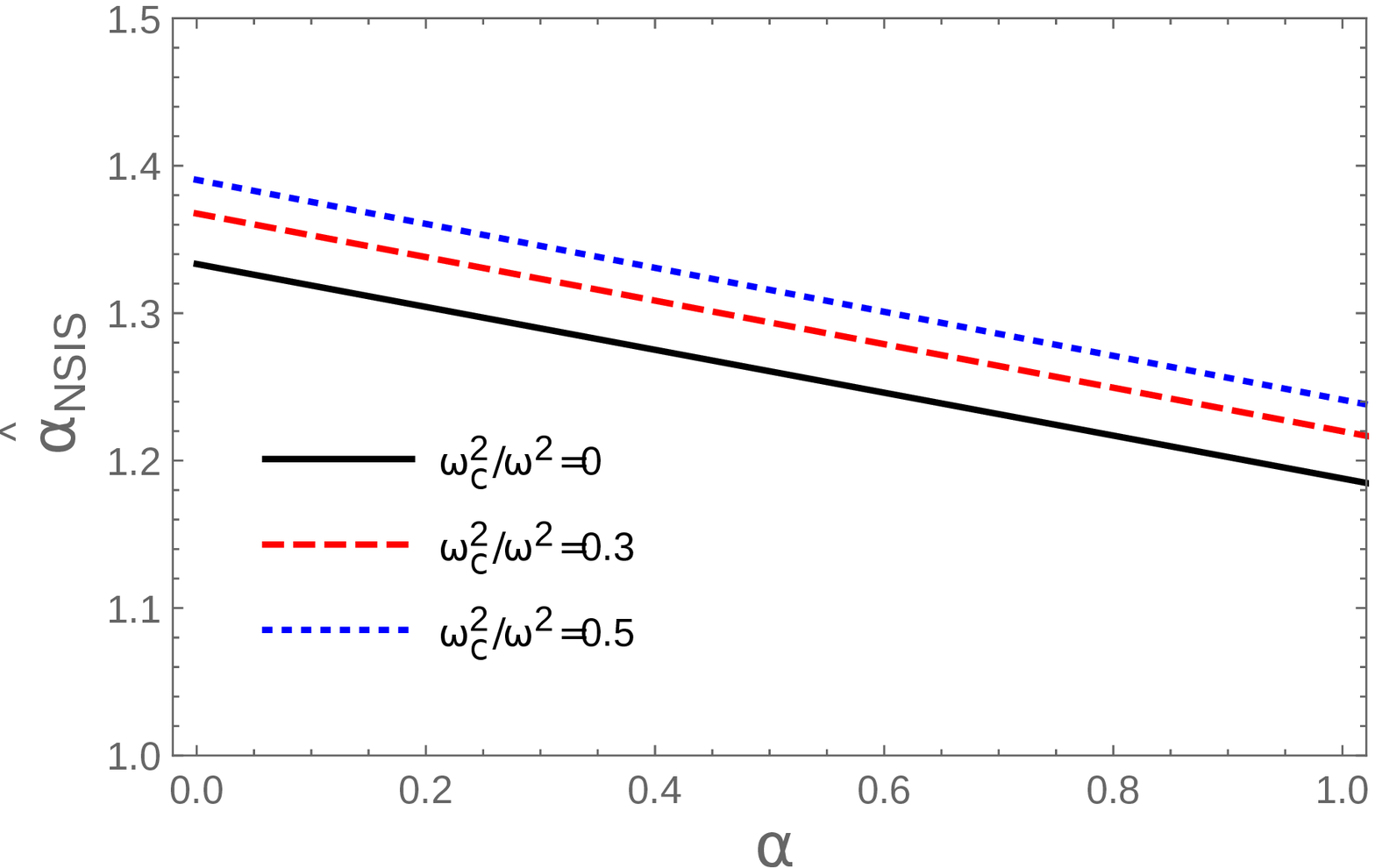}
  \end{center}
\caption{Plot of deflection angle $\hat{\alpha}_\mathrm{NSIS}$ as a function of $\frac{\omega^2_{c}}{\omega^2}$
(upper panel) and $\alpha$ (lower panel). The fixed parameters used
are $b=3$ and $r_c=3$.}\label{angleNsis2}
\end{figure}

\begin{figure}[h!]
 \begin{center}
   \includegraphics[scale=0.38]{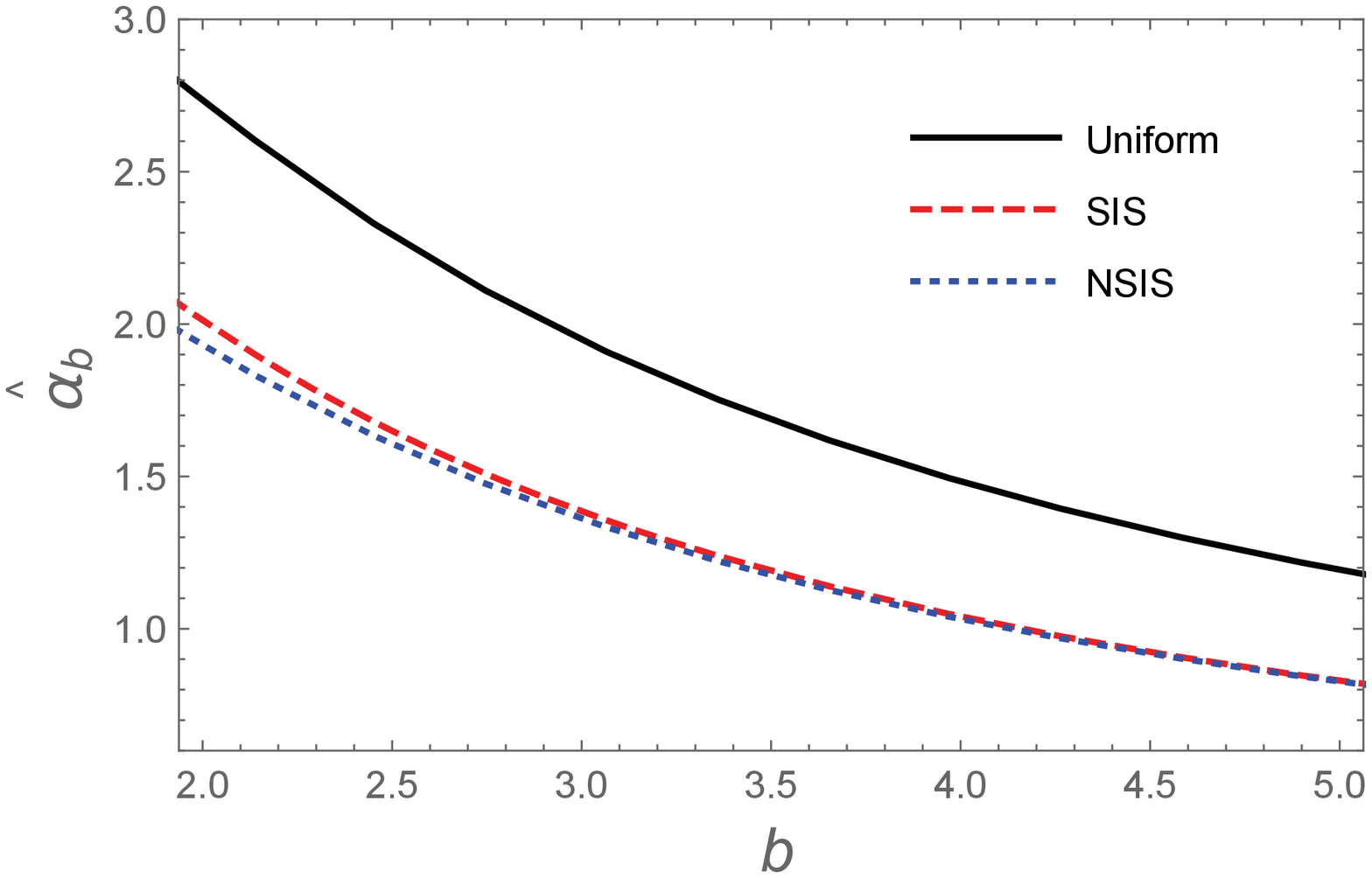}
   \includegraphics[scale=0.38]{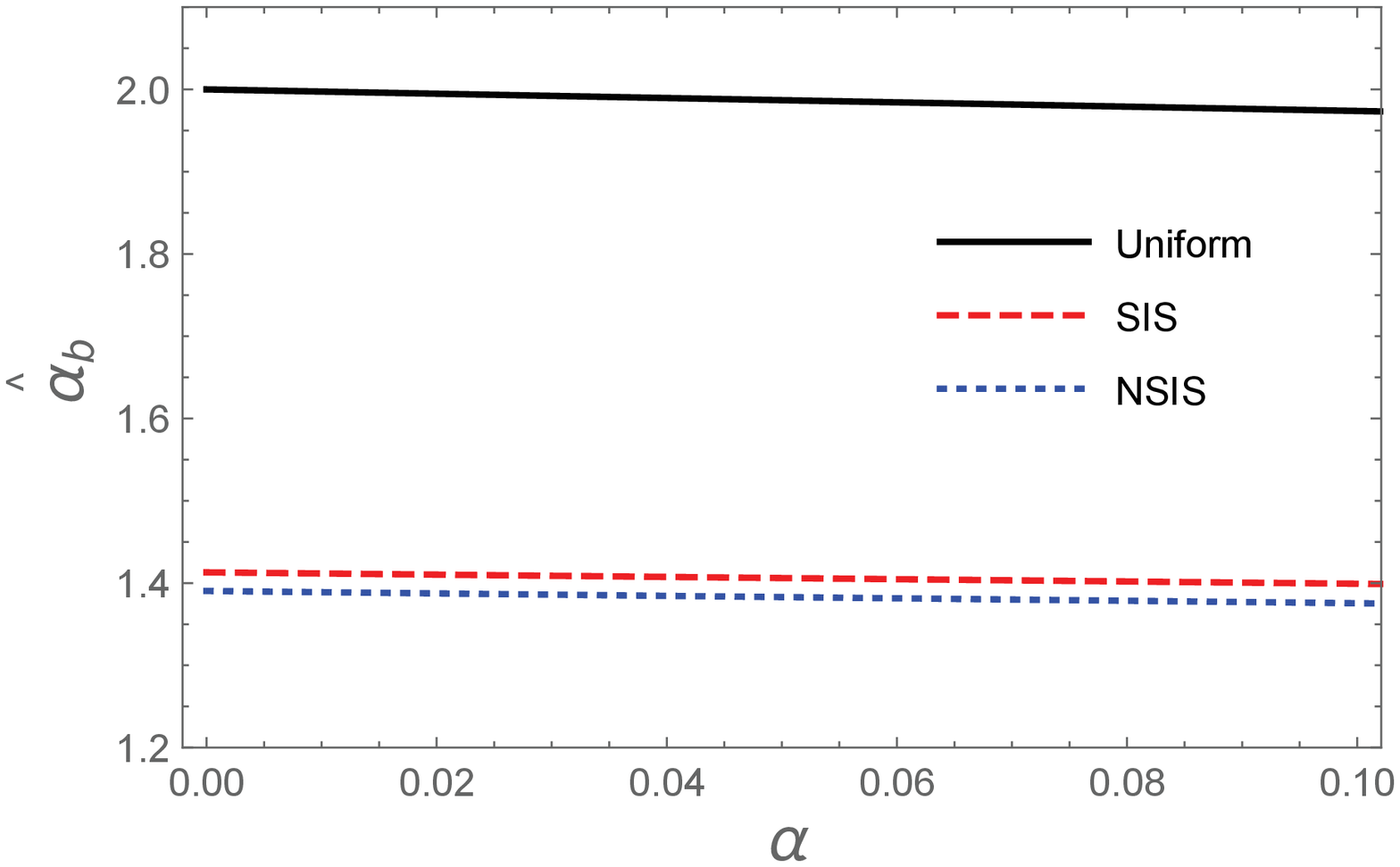}
  \end{center}
\caption{Plot of the deflection angle $\hat{\alpha}_{b}$ as a function of the impact
parameter $b$ (upper panel) and $\alpha$ (lower panel). The corresponding fixed parameters used are
$\frac{\omega^2_{0}}{\omega^2}=0.5$, $\frac{\omega^2_{c}}{\omega^2}=0.5$, $b=3$ and $r_c=3$.}\label{angle}
\end{figure}

As previously executed we employ the same graphical interpretation to unfold the properties of NSIS concerning the photon motion.
Here, the distribution of NSIS is associated with the parameter $\frac{\omega^2_{c}}{\omega^2}$. It is evidently revealed
from Fig. (\ref{angleNsis1},\ref{angleNsis2}) that the behaviour of the impact parameter $b$,
coupling constant $\alpha$ and the parameter $\frac{\omega^2_{c}}{\omega^2}$
cannot be distinguished from a specific point of view when brought in comparison with the uniform plasma and SIS case.
Nevertheless, one can at least figure out the distribution which has the most pronounced effect on the deflection angle.
Fig. (\ref{angle}) is a visual juxtaposition of the $\hat{\alpha}_\mathrm{uni}$, $\hat{\alpha}_\mathrm{SIS}$ and $\hat{\alpha}_\mathrm{NSIS}$
as a function of the impact parameter and the coupling constant. It is quite obvious that the deflection is maximum
when the black hole is surrounded by a uniform plasma medium. The final result can therefore be encaspulated in a mathematical expression as,
$\hat{\alpha}_\mathrm{uni}> \hat{\alpha}_\mathrm{SIS}>\hat{\alpha}_\mathrm{NSIS}$.

\section{Lens equation and magnification in the presence of plasma}\label{Magnification}
We now focus on the magnification of the image source brightness in
the presence of plasma using the angle of deflection $\hat{\alpha}$
discussed in our previous sections, particulary for the uniform plasma and SIS case. Fig. (\ref{Lensingsystem}) is a formal
visualization of the gravitational lensing system where the black hole
function as a lens to the distant source (star) and generates an image. In the diagram
$D_\mathrm{s}$, $D_\mathrm{d}$, and $D_\mathrm{ds}$ represent the distances from the source to the
observer, from the lens to the observer, and from the source to the lens, respectively. The angular
position of the source and image are correspondingly specified by $\beta$ and $\theta$.
The relation which finds the angular position
of the image, usually termed as the gravitational lens equation is defined as follows \cite{Abu:2013a,Abu:2017a}

\begin{align}\label{lenseq}
\theta D_\mathrm{s}=\beta D_\mathrm{s}+\hat{\alpha}D_\mathrm{ds}.
\end{align}
Generally, in this setup the impact parameter $b$ admits the expression $b= D_\mathrm{d}\theta$ which on substitution in (\ref{lenseq})
along with a newly introduced quantity $\xi(\theta)=|\hat{\alpha}_b|b$ yields the following equation
\begin{align}\label{newlenseq}
\beta=\theta -\frac{D_\mathrm{ds}}{D_\mathrm{s}}\frac{\xi(\theta)}{D_\mathrm{d}}\frac{1}{\theta}.
\end{align}
By making a special choice of $\beta=0$ one may obtain a uniquely aligned
or more precisely a collinear framework of the observer, the lens and the source.
In such context due to symmetry around the lens the image appears in the form of a ring, often called the Einstein ring
with a radius $R_0=D_\mathrm{d}\theta_0$. Here, $\theta_0$ is referred as the Einstein angle, which for
the Schwarzschild space-time becomes

\begin{align}
\theta_0=\sqrt{2R_s\frac{D_{ds}}{D_dD_s}}.
\end{align}

The lensing process by the astrophysical objects is trackable only through the
apparent brightness of the source which can be computed utilizing the formula
\begin{align}\label{magni}
\mu_{\Sigma}=\frac{I_\mathrm{tot}}{I_*}=\underset{k}\sum\bigg|\bigg(\frac{\theta_k}{\beta}\bigg)\bigg(\frac{d\theta_k}{d\beta}\bigg)\bigg|, \quad k=1,2,.....,j,
\end{align}
where $k$ denotes the number of images formed each indicated by the angular position $\theta_k$, $I_\mathrm{tot}$ is the total brightness of all the images
and $I_*$ is the unlensed brightness of the source. However, this
straightforward relation cannot be efficiently employed for a computation
of the total magnification, therefore we follow the notion in \cite{Schnei:1999a}
by using (\ref{magni}) in terms of the positive- and negative-parity images. As a customary, the positive
parity image lies on the same side of the lens as the source for $\beta>0$, whereas the negative-parity image lies on the opposite side of the lens from
the source for $\beta<0$. Thus, the verified formulae for the magnification of the source given in \cite{Schnei:1999a} are as follows

\begin{align}\label{postparity}
\mu^\mathrm{pl}_\mathrm{+}=\frac{1}{4}\bigg(\frac{x}{\sqrt{x^2+4}}+\frac{\sqrt{x^2+4}}{x}+2\bigg),
\end{align}

\begin{align}\label{negparity}
\mu^\mathrm{pl}_\mathrm{-}=\frac{1}{4}\bigg(\frac{x}{\sqrt{x^2+4}}+\frac{\sqrt{x^2+4}}{x}-2\bigg),
\end{align}
where $x=\frac{\beta}{\theta_0}$ is a dimensionless quantity and the superscript $\mathrm{pl}$ refers to the plasma presence. After performing
some calculations with the aid of (\ref{postparity},\ref{negparity})
we get the total magnification of the image as below

\begin{align}\label{magtot}
\mu^\mathrm{pl}_\mathrm{tot}=\mu^\mathrm{pl}_{+}+\mu^\mathrm{pl}_{-}=\frac{x^2+2}{x\sqrt{x^2+4}}.
\end{align}
Henceforward, we shall concentrate on the uniform plasma and SIS case to examine
the magnification of the apparent brightness of the source.

\subsection{$\mathbf{Uniform}$ $\mathbf{Plasma}$}
\begin{figure}[h!]
 \begin{center}
   \includegraphics[align=t,scale=0.4]{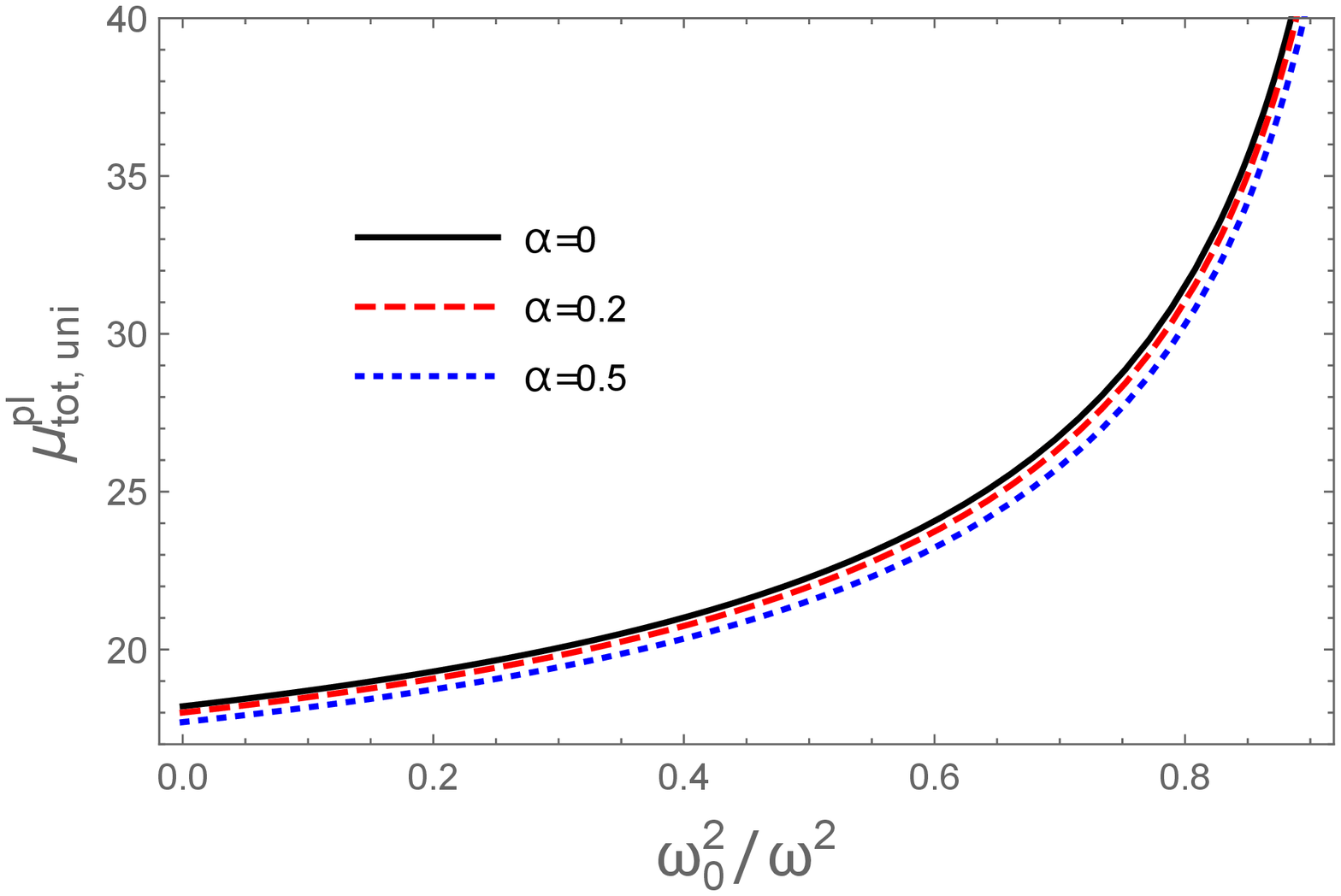}
    \includegraphics[align=t,scale=0.4]{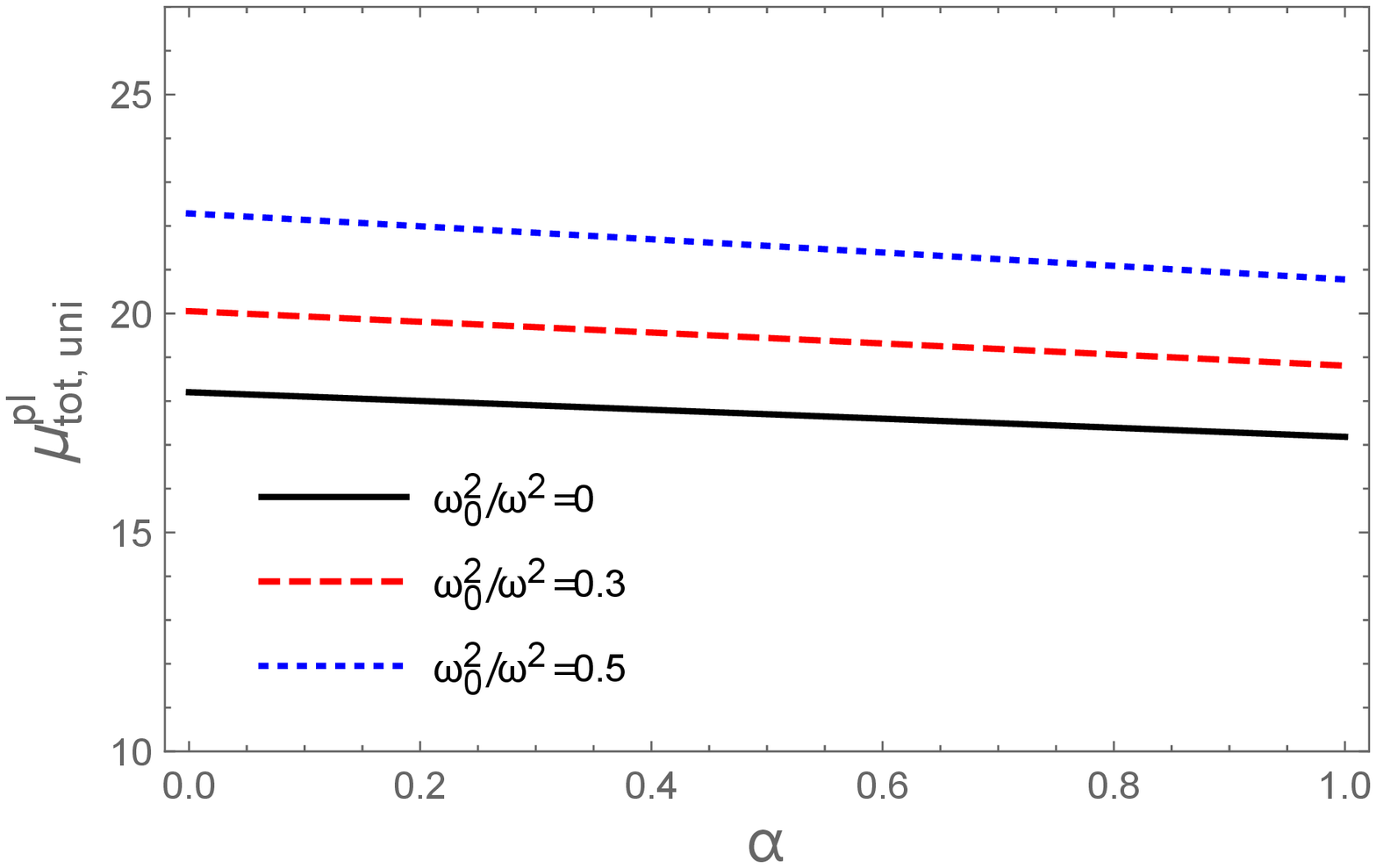}
  \end{center}
\caption{The total magnification of the image brightness in the presence of uniform plasma as a function of $\frac{\omega^2_0}{\omega^2}$
(upper panel) and $\alpha$ (lower panel). The fixed parameters
used are $R_s=2$, $b=3$ and $x_0=0.055$.}\label{maguni}
\end{figure}

\begin{figure}[h!]
 \begin{center}
   \includegraphics[align=t,scale=0.4]{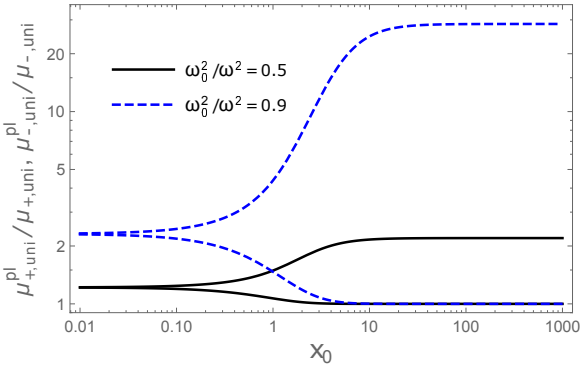}
  \end{center}
\caption{Plots for $\mu^\mathrm{pl}_\mathrm{+,uni}/\mu_\mathrm{+,uni}$ (lower curve) and $\mu^\mathrm{pl}_\mathrm{-,uni}/\mu_\mathrm{-,uni}$ (upper curve)
of the image magnifications in the presence of uniform plasma. The fixed parameters used are $R_s=2$, $b=3$, $\alpha=0.5$
and $x_0=0.055$.}\label{magratiouni}
\end{figure}

Keeping in view the discussion regarding  the 4D-EGB gravity surrounded by a uniform plasma
the angular position of the image is smoothly worked out from (\ref{unialpha},\ref{newlenseq}) as below,

\begin{align}\label{anguniplasma}
\theta^{\mathrm{pl}}_{\mathrm{uni}}=\theta_0\sqrt{\frac{1}{2}\bigg(\bigg(1-\frac{3\pi \alpha R_s}{16b^3}\bigg)+
\bigg(1-\frac{3\pi \alpha R_s}{4b^3}\bigg)\frac{1}{\big(1-\frac{\omega^2_0}{\omega^2}\big)}\bigg)}.
\end{align}
The value of $x$ requisite for the magnification of the source in uniform plasma thus takes the form

\begin{align}
x=\frac{x_0}{\sqrt{\frac{1}{2}\bigg(\bigg(1-\frac{3\pi \alpha R_s}{16b^3}\bigg)+
\bigg(1-\frac{3\pi \alpha R_s}{4b^3}\bigg)\frac{1}{\big(1-\frac{\omega^2_0}{\omega^2}\big)}\bigg)}},
\end{align}
where $x_0=\beta/\theta_0$. In the upper panel of Fig. (\ref{maguni}) the total magnification is plotted against
the plasma parameter $\frac{\omega^2_0}{\omega^2}$ for various coupling parameters $\alpha$. The magnification is observed to increase for
increasing uniform plasma distribution. Hence, we restore the fact that the presence of uniform plasma typically increases the magnification and immensely reduces in case of the vacuum\cite{Bin:2010a}. In the lower panel, for different values of $\frac{\omega^2_0}{\omega^2}$ the behaviour of  $\mu^\mathrm{pl}_\mathrm{tot,uni}$ is shown by varying $\alpha$. Generally, for increasing $\alpha$, $\mu^\mathrm{pl}_\mathrm{tot,uni}$ decreases and is maximum for the Schwarzschild back hole ($\alpha=0$). For the purpose of additional details we hereby discuss the magnification ratio of uniform plasma to vacuum assuming the positive- and negative- parity images, see Fig.(\ref{magratiouni}).
The lower and upper curves depict the plots for $\mu^\mathrm{pl}_\mathrm{+,uni}/\mu_\mathrm{+,uni}$ and
$\mu^\mathrm{pl}_\mathrm{-,uni}/\mu_\mathrm{-,uni}$, respectively. It is observed that the two curves meet at a single point each time $x_0 \rightarrow 0$,
for $\frac{\omega^2_0}{\omega^2}$=0.5 and 0.9 the corresponding values of $x_0$ recorded are 1.21766 and 2.31176. On the whole, at large $x_0$ the ratio
$\mu^\mathrm{pl}_\mathrm{+,uni}/\mu_\mathrm{+,uni}$ tends to unity for each $\frac{\omega^2_0}{\omega^2}$,
but exclusive for $\mu^\mathrm{pl}_\mathrm{-,uni}/\mu_\mathrm{-,uni}$ the magnification factor is greatly amplified for a larger plasma distribution.

\subsection{$\mathbf{Singular}$ $\mathbf{Isothermal}$ $\mathbf{sphere}$}
We shall revisit the above analysis in a similar manner to discover the influence of SIS on magnification of the image source.
Hence, the angular position of the image in this background is obtained from (\ref{alphasis},\ref{newlenseq})
as below,

\begin{figure}[h!]
 \begin{center}
   \includegraphics[align=t,scale=0.4]{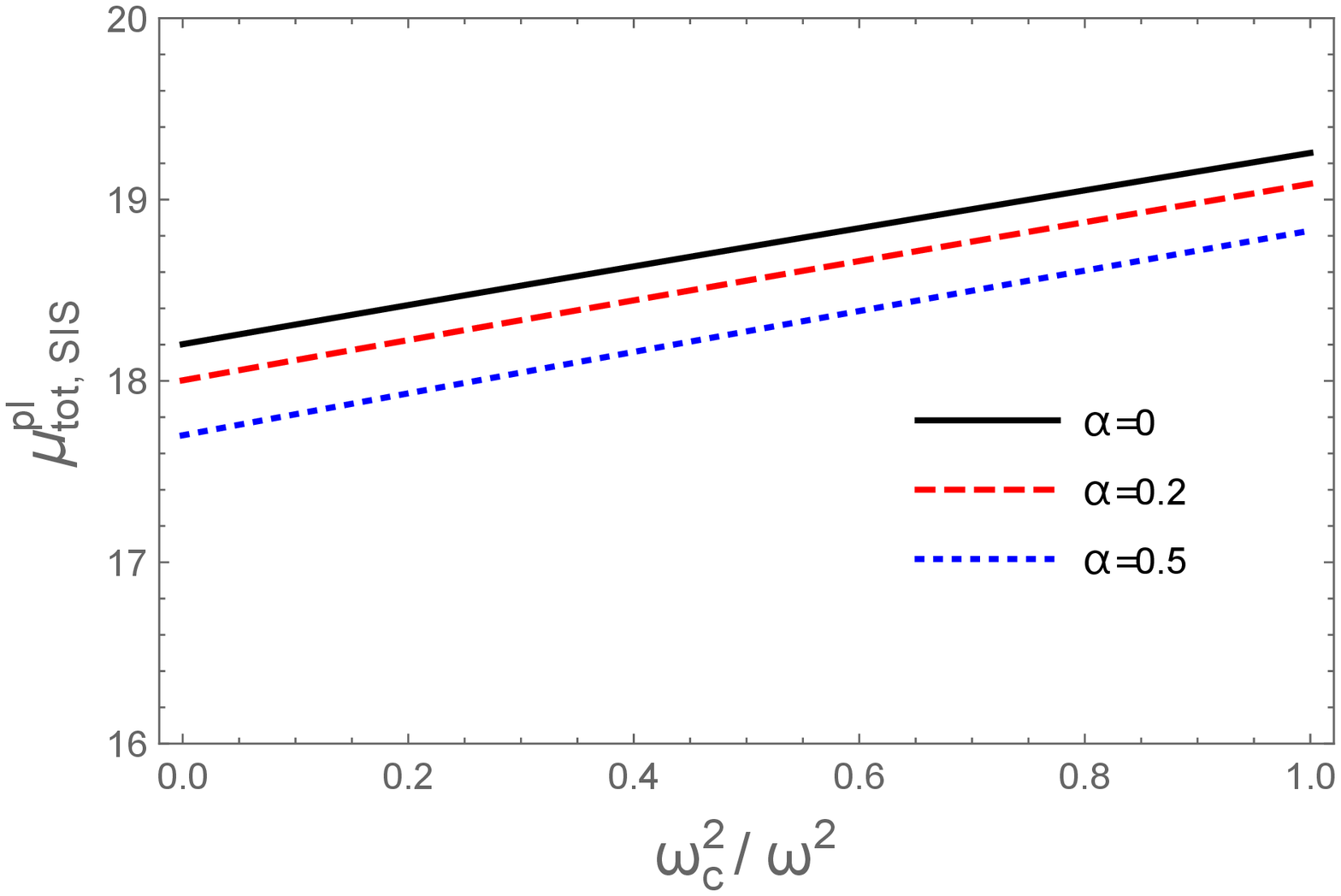}
   \includegraphics[align=t,scale=0.4]{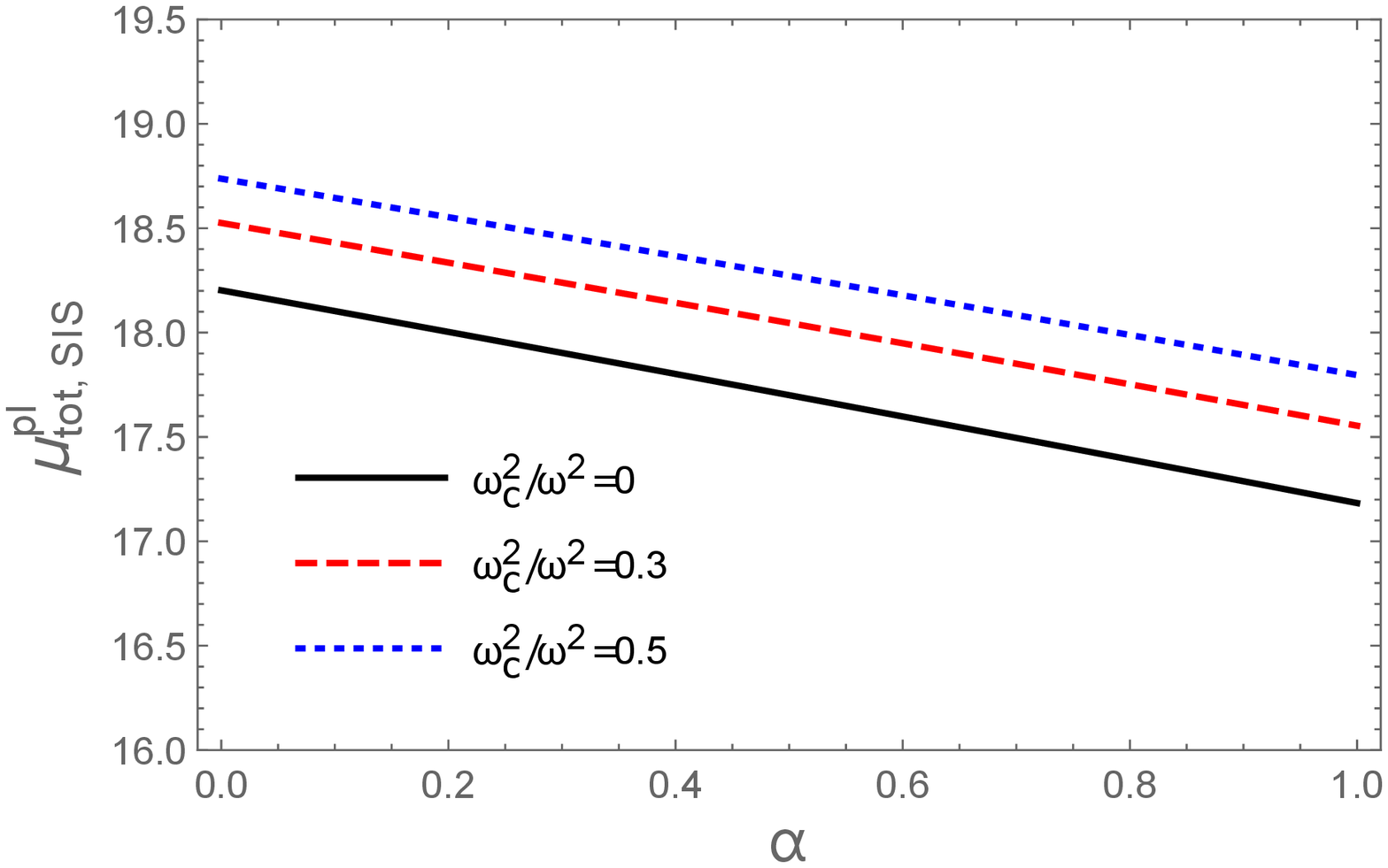}
  \end{center}
\caption{The total magnification of the image brightness in the presence of SIS as a function of $\frac{\omega^2_c}{\omega^2}$
(upper panel) and $\alpha$ (lower panel). The fixed parameters
used are $R_s=2$, $b=3$ and $x_0=0.055$.}\label{magsis}
\end{figure}

\begin{figure}[h!]
 \begin{center}
   \includegraphics[align=t,scale=0.4]{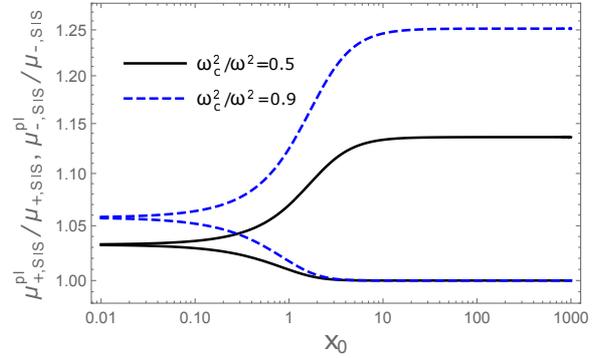}
  \end{center}
\caption{Plots for $\mu^\mathrm{pl}_\mathrm{+,SIS}/\mu_\mathrm{+,SIS}$ (lower curve) and $\mu^\mathrm{pl}_\mathrm{-,SIS}/\mu_\mathrm{-,SIS}$ (upper curve)
of the image magnifications in the presence of SIS. The fixed parameters used are $R_s=2$,
$b=3$, $\alpha=0.5$ and $x_0=0.055$.}\label{magsisplasma}
\end{figure}

\begin{figure}[h!]
 \begin{center}
   \includegraphics[scale=0.4]{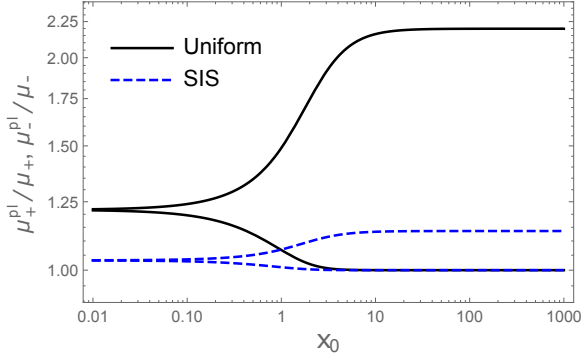}
  \end{center}
\caption{ Plots for $\mu^\mathrm{pl}_\mathrm{+}/\mu_\mathrm{+}$ (lower curve) and $\mu^\mathrm{pl}_\mathrm{-}/\mu_\mathrm{-}$ (upper curve)
of the image magnifications in the presence of uniform plasma and SIS. The fixed parameters used are $R_s=2$, $b=3$, $\frac{\omega^2_0}{\omega^2}=0.5$,
$\frac{\omega^2_c}{\omega^2}=0.5$, $\alpha=0.5$ and $x_0=0.055$.}\label{magus}
\end{figure}

\begin{align}
\theta^\mathrm{pl}_\mathrm{SIS}=\theta_0\sqrt{\frac{1}{2}\bigg(2-\frac{15\pi \alpha R_s}{16b^3}+\frac{R_s\omega^2_c}{b\omega^2}\bigg(
\frac{1}{2}-\frac{2R_s}{3b\pi}+\frac{5\alpha R^2_s}{8b^4}\bigg)\bigg)}.
\end{align}
The value of $x$ for the magnification of the source in the SIS surroundings is thus expressed in the form
\begin{align}
x=\frac{x_0}{\sqrt{\frac{1}{2}\bigg(2-\frac{15\pi \alpha R_s}
{16b^3}+\frac{R_s\omega^2_c}{b\omega^2}\bigg(\frac{1}{2}-\frac{2R_s}{3b\pi}+\frac{5\alpha R^2_s}{8b^4}\bigg)\bigg)}},
\end{align}
where $x_0=\frac{\beta}{\theta_0}$. The behaviour of total magnification of the image source dependence of the SIS distribution $\frac{\omega^2_c}{\omega^2}$
for  distinct $\alpha$ is displayed in the upper panel of Fig. (\ref{magsis}). It is verified that the presence
of SIS amplifies the luminosity of the source star.
Alternatively, in the lower panel by increasing the $\alpha$ parameter a decrease is observed for $\mu^\mathrm{pl}_\mathrm{tot,SIS}$.
In Fig. (\ref{magsisplasma}), the magnification ratio of SIS to vacuum operates analogous to the uniform
plasma case when adjusted with the same parameters, but nonetheless, when the black hole is surrounded by a uniform plasma medium,
a considerable magnification is perceived by a far-off observer ( see Fig. (\ref{magus}))

\section{Conclusion} \label{con}
We discussed categorically the weak-field lensing phenomenon in the background of a 4 dimensional
Einstein-Gauss-Bonnet gravity environed by a uniform plasma, singular isothermal sphere and non-singular isothermal sphere.
The deflection angle for each case is incisively discussed to evince the most effective medium. Interestingly, the
uniform plasma is investigated to deviate the incoming photons at a very large angle.
Moreover, the SIS deflection angle is greater to a certain extent
than that of the NSIS, even though the difference could be regarded as negligible.
In general, the light lensed by a nearby black hole modifies the spectrum of the source star that results in the apparent
magnification of the image source luminosity. This behaviour has remained the focus of our analysis by considering
a plasma environment and we have successfully achieved the fact that the presence of a uniform plasma significantly enhances the magnification.
Finally, all along the examination, the Schwarzschild black hole compared to the 4D-EGB black hole emerged as an effective gravity.

\section{Acknowledgement} F.A. acknowledges the support of INHA
University in Tashkent.

\bibliographystyle{apsrev4-1}
\bibliography{Lensing}

%merlin.mbs apsrev4-1.bst 2010-07-25 4.21a (PWD, AO, DPC) hacked
%Control: key (0)
%Control: author (72) initials jnrlst
%Control: editor formatted (1) identically to author
%Control: production of article title (-1) disabled
%Control: page (0) single
%Control: year (1) truncated
%Control: production of eprint (0) enabled
\begin{thebibliography}{80}%
\makeatletter
\providecommand \@ifxundefined [1]{%
 \@ifx{#1\undefined}
}%
\providecommand \@ifnum [1]{%
 \ifnum #1\expandafter \@firstoftwo
 \else \expandafter \@secondoftwo
 \fi
}%
\providecommand \@ifx [1]{%
 \ifx #1\expandafter \@firstoftwo
 \else \expandafter \@secondoftwo
 \fi
}%
\providecommand \natexlab [1]{#1}%
\providecommand \enquote  [1]{``#1''}%
\providecommand \bibnamefont  [1]{#1}%
\providecommand \bibfnamefont [1]{#1}%
\providecommand \citenamefont [1]{#1}%
\providecommand \href@noop [0]{\@secondoftwo}%
\providecommand \href [0]{\begingroup \@sanitize@url \@href}%
\providecommand \@href[1]{\@@startlink{#1}\@@href}%
\providecommand \@@href[1]{\endgroup#1\@@endlink}%
\providecommand \@sanitize@url [0]{\catcode `\\12\catcode `\$12\catcode
  `\&12\catcode `\#12\catcode `\^12\catcode `\_12\catcode `\%12\relax}%
\providecommand \@@startlink[1]{}%
\providecommand \@@endlink[0]{}%
\providecommand \url  [0]{\begingroup\@sanitize@url \@url }%
\providecommand \@url [1]{\endgroup\@href {#1}{\urlprefix }}%
\providecommand \urlprefix  [0]{URL }%
\providecommand \Eprint [0]{\href }%
\providecommand \doibase [0]{http://dx.doi.org/}%
\providecommand \selectlanguage [0]{\@gobble}%
\providecommand \bibinfo  [0]{\@secondoftwo}%
\providecommand \bibfield  [0]{\@secondoftwo}%
\providecommand \translation [1]{[#1]}%
\providecommand \BibitemOpen [0]{}%
\providecommand \bibitemStop [0]{}%
\providecommand \bibitemNoStop [0]{.\EOS\space}%
\providecommand \EOS [0]{\spacefactor3000\relax}%
\providecommand \BibitemShut  [1]{\csname bibitem#1\endcsname}%
\let\auto@bib@innerbib\@empty
%</preamble>
\bibitem [{\citenamefont {Synge}()}]{Synge:1960b}%
  \BibitemOpen
  \bibfield  {author} {\bibinfo {author} {\bibfnamefont {J.~L.}\ \bibnamefont
  {Synge}},\ }\href@noop {} {\bibinfo  {journal} {Relativity: The General
  Theory. North-Holland, Amsterdam, 1960}\ }\BibitemShut {NoStop}%
\bibitem [{\citenamefont {Schneider}\ \emph {et~al.}()\citenamefont
  {Schneider}, \citenamefont {J.Ehlers},\ and\ \citenamefont
  {Falco}}]{Schnei:1999a}%
  \BibitemOpen
\bibfield  {journal} {  }\bibfield  {author} {\bibinfo {author} {\bibfnamefont
  {P.}~\bibnamefont {Schneider}}, \bibinfo {author} {\bibnamefont {J.Ehlers}},
  \ and\ \bibinfo {author} {\bibfnamefont {E.}~\bibnamefont {Falco}},\
  }\href@noop {} {\bibinfo  {journal} {Gravitational Lenses. Astronomy and
  Astrophysics Library (Springer, 1999). ISSN 0941-7834}\ }\BibitemShut
  {NoStop}%
\bibitem [{\citenamefont {Perlick}()}]{Perlick:2000a}%
  \BibitemOpen
\bibfield  {journal} {  }\bibfield  {author} {\bibinfo {author} {\bibfnamefont
  {V.}~\bibnamefont {Perlick}},\ }\href@noop {} {\bibinfo  {journal} {Ray
  Optics, Fermat’s Principle, and Applications to General Relativity,
  (Springer, Berlin, 2000)}\ }\BibitemShut {NoStop}%
\bibitem [{\citenamefont {Perlick}(2004)}]{Perlick:2004a}%
  \BibitemOpen
\bibfield  {journal} {  }\bibfield  {author} {\bibinfo {author} {\bibfnamefont
  {V.}~\bibnamefont {Perlick}},\ }\href {\doibase 10.12942/lrr-2004-9}
  {\bibfield  {journal} {\bibinfo  {journal} {Living Rev.~Relativ.}\ }\textbf
  {\bibinfo {volume} {7}},\ \bibinfo {pages} {9} (\bibinfo {year}
  {2004})}\BibitemShut {NoStop}%
\bibitem [{\citenamefont {Virbhadra}\ and\ \citenamefont
  {Ellis}(2000)}]{Virbha:2000a}%
  \BibitemOpen
  \bibfield  {author} {\bibinfo {author} {\bibfnamefont {K.~S.}\ \bibnamefont
  {Virbhadra}}\ and\ \bibinfo {author} {\bibfnamefont {G.~F.~R.}\ \bibnamefont
  {Ellis}},\ }\href {\doibase 10.1103/PhysRevD.62.084003} {\bibfield  {journal}
  {\bibinfo  {journal} {Phys.~Rev.~D.}\ }\textbf {\bibinfo {volume} {62}},\
  \bibinfo {pages} {084003} (\bibinfo {year} {2000})}\BibitemShut {NoStop}%
\bibitem [{\citenamefont {Virbhadra}\ and\ \citenamefont
  {Ellis}(2002)}]{Virbha:2002a}%
  \BibitemOpen
  \bibfield  {author} {\bibinfo {author} {\bibfnamefont {K.~S.}\ \bibnamefont
  {Virbhadra}}\ and\ \bibinfo {author} {\bibfnamefont {G.~F.~R.}\ \bibnamefont
  {Ellis}},\ }\href {\doibase 10.1103/PhysRevD.65.103004} {\bibfield  {journal}
  {\bibinfo  {journal} {Phys.~Rev.~D.}\ }\textbf {\bibinfo {volume} {65}},\
  \bibinfo {pages} {103004} (\bibinfo {year} {2002})}\BibitemShut {NoStop}%
\bibitem [{\citenamefont {{Bozza}}\ \emph {et~al.}(2001)\citenamefont
  {{Bozza}}, \citenamefont {{Capozziello}}, \citenamefont {{Iovane}},\ and\
  \citenamefont {{Scarpetta}}}]{Bozza:2001a}%
  \BibitemOpen
  \bibfield  {author} {\bibinfo {author} {\bibfnamefont {V.}~\bibnamefont
  {{Bozza}}}, \bibinfo {author} {\bibfnamefont {S.}~\bibnamefont
  {{Capozziello}}}, \bibinfo {author} {\bibfnamefont {G.}~\bibnamefont
  {{Iovane}}}, \ and\ \bibinfo {author} {\bibfnamefont {G.}~\bibnamefont
  {{Scarpetta}}},\ }\href {\doibase 10.1023/A:1012292927358} {\bibfield
  {journal} {\bibinfo  {journal} {Gen. Rel. Grav.}\ }\textbf {\bibinfo {volume}
  {33}},\ \bibinfo {pages} {1535} (\bibinfo {year} {2001})},\ \Eprint
  {http://arxiv.org/abs/gr-qc/0102068} {arXiv:gr-qc/0102068 [gr-qc]}
  \BibitemShut {NoStop}%
\bibitem [{\citenamefont {{Bozza}}(2002)}]{Bozza:2002b}%
  \BibitemOpen
  \bibfield  {author} {\bibinfo {author} {\bibfnamefont {V.}~\bibnamefont
  {{Bozza}}},\ }\href {\doibase 10.1103/PhysRevD.66.103001} {\bibfield
  {journal} {\bibinfo  {journal} {Phys.~Rev.~D.}\ }\textbf {\bibinfo {volume}
  {66}},\ \bibinfo {eid} {103001} (\bibinfo {year} {2002})},\ \Eprint
  {http://arxiv.org/abs/gr-qc/0208075} {arXiv:gr-qc/0208075 [gr-qc]}
  \BibitemShut {NoStop}%
\bibitem [{\citenamefont {{Bozza}}(2003)}]{Bozza:2003a}%
  \BibitemOpen
  \bibfield  {author} {\bibinfo {author} {\bibfnamefont {V.}~\bibnamefont
  {{Bozza}}},\ }\href {\doibase 10.1103/PhysRevD.67.103006} {\bibfield
  {journal} {\bibinfo  {journal} {Phys.~Rev.~D.}\ }\textbf {\bibinfo {volume}
  {67}},\ \bibinfo {eid} {103006} (\bibinfo {year} {2003})},\ \Eprint
  {http://arxiv.org/abs/gr-qc/0210109} {arXiv:gr-qc/0210109 [gr-qc]}
  \BibitemShut {NoStop}%
\bibitem [{\citenamefont {S.~E.~V$\acute{a}$zquez}(2004)}]{Sevb:2004a}%
  \BibitemOpen
  \bibfield  {author} {\bibinfo {author} {\bibfnamefont {E.~P.~E.}\
  \bibnamefont {S.~E.~V$\acute{a}$zquez}},\ }\href {\doibase
  10.1393/ncb/i2004-10121-y} {\bibfield  {journal} {\bibinfo  {journal} {Nuovo
  Cim.~B}\ }\textbf {\bibinfo {volume} {119}},\ \bibinfo {pages} {489}
  (\bibinfo {year} {2004})}\BibitemShut {NoStop}%
\bibitem [{\citenamefont {{Bozza}}\ \emph {et~al.}(2005)\citenamefont
  {{Bozza}}, \citenamefont {{de Luca}}, \citenamefont {{Scarpetta}},\ and\
  \citenamefont {{Sereno}}}]{Bozza:2005a}%
  \BibitemOpen
  \bibfield  {author} {\bibinfo {author} {\bibfnamefont {V.}~\bibnamefont
  {{Bozza}}}, \bibinfo {author} {\bibfnamefont {F.}~\bibnamefont {{de Luca}}},
  \bibinfo {author} {\bibfnamefont {G.}~\bibnamefont {{Scarpetta}}}, \ and\
  \bibinfo {author} {\bibfnamefont {M.}~\bibnamefont {{Sereno}}},\ }\href
  {\doibase 10.1103/PhysRevD.72.083003} {\bibfield  {journal} {\bibinfo
  {journal} {Phys.~Rev.~D.}\ }\textbf {\bibinfo {volume} {72}},\ \bibinfo {eid}
  {083003} (\bibinfo {year} {2005})},\ \Eprint
  {http://arxiv.org/abs/gr-qc/0507137} {arXiv:gr-qc/0507137 [gr-qc]}
  \BibitemShut {NoStop}%
\bibitem [{\citenamefont {{Bozza}}\ \emph {et~al.}(2006)\citenamefont
  {{Bozza}}, \citenamefont {{de Luca}},\ and\ \citenamefont
  {{Scarpetta}}}]{Bozza:2006a}%
  \BibitemOpen
  \bibfield  {author} {\bibinfo {author} {\bibfnamefont {V.}~\bibnamefont
  {{Bozza}}}, \bibinfo {author} {\bibfnamefont {F.}~\bibnamefont {{de Luca}}},
  \ and\ \bibinfo {author} {\bibfnamefont {G.}~\bibnamefont {{Scarpetta}}},\
  }\href {\doibase 10.1103/PhysRevD.74.063001} {\bibfield  {journal} {\bibinfo
  {journal} {Phys.~Rev.~D.}\ }\textbf {\bibinfo {volume} {74}},\ \bibinfo {eid}
  {063001} (\bibinfo {year} {2006})},\ \Eprint
  {http://arxiv.org/abs/gr-qc/0604093} {arXiv:gr-qc/0604093 [gr-qc]}
  \BibitemShut {NoStop}%
\bibitem [{\citenamefont {Eiroa}\ \emph {et~al.}(2002)\citenamefont {Eiroa},
  \citenamefont {Romero},\ and\ \citenamefont {Torres}}]{Eiroa:2002b}%
  \BibitemOpen
  \bibfield  {author} {\bibinfo {author} {\bibfnamefont {E.~F.}\ \bibnamefont
  {Eiroa}}, \bibinfo {author} {\bibfnamefont {G.~E.}\ \bibnamefont {Romero}}, \
  and\ \bibinfo {author} {\bibfnamefont {D.~F.}\ \bibnamefont {Torres}},\
  }\href {\doibase 10.1103/PhysRevD.66.024010} {\bibfield  {journal} {\bibinfo
  {journal} {Phys.~Rev.~D.}\ }\textbf {\bibinfo {volume} {66}},\ \bibinfo
  {pages} {024010} (\bibinfo {year} {2002})}\BibitemShut {NoStop}%
\bibitem [{\citenamefont {Eiroa}\ and\ \citenamefont
  {Torres}(2004)}]{Eiroa:2004a}%
  \BibitemOpen
  \bibfield  {author} {\bibinfo {author} {\bibfnamefont {E.~F.}\ \bibnamefont
  {Eiroa}}\ and\ \bibinfo {author} {\bibfnamefont {D.~F.}\ \bibnamefont
  {Torres}},\ }\href {\doibase 10.1103/PhysRevD.69.063004} {\bibfield
  {journal} {\bibinfo  {journal} {Phys.~Rev.~D.}\ }\textbf {\bibinfo {volume}
  {69}},\ \bibinfo {pages} {063004} (\bibinfo {year} {2004})}\BibitemShut
  {NoStop}%
\bibitem [{\citenamefont {Eiroa}(2005)}]{Eiroa:2005a}%
  \BibitemOpen
  \bibfield  {author} {\bibinfo {author} {\bibfnamefont {E.~F.}\ \bibnamefont
  {Eiroa}},\ }\href {\doibase 10.1103/PhysRevD.71.083010} {\bibfield  {journal}
  {\bibinfo  {journal} {Phys.~Rev.~D.}\ }\textbf {\bibinfo {volume} {71}},\
  \bibinfo {pages} {083010} (\bibinfo {year} {2005})}\BibitemShut {NoStop}%
\bibitem [{\citenamefont {Virbhadra}(2009)}]{Virbha:2009b}%
  \BibitemOpen
  \bibfield  {author} {\bibinfo {author} {\bibfnamefont {K.~S.}\ \bibnamefont
  {Virbhadra}},\ }\href {\doibase 10.1103/PhysRevD.79.083004} {\bibfield
  {journal} {\bibinfo  {journal} {Phys.~Rev.~D.}\ }\textbf {\bibinfo {volume}
  {79}},\ \bibinfo {pages} {083004} (\bibinfo {year} {2009})}\BibitemShut
  {NoStop}%
\bibitem [{\citenamefont {Wei}\ \emph {et~al.}(2012)\citenamefont {Wei},
  \citenamefont {Liu}, \citenamefont {Fu},\ and\ \citenamefont
  {Yang}}]{Wei:2012b}%
  \BibitemOpen
  \bibfield  {author} {\bibinfo {author} {\bibfnamefont {S.-W.}\ \bibnamefont
  {Wei}}, \bibinfo {author} {\bibfnamefont {Y.-X.}\ \bibnamefont {Liu}},
  \bibinfo {author} {\bibfnamefont {C.-E.}\ \bibnamefont {Fu}}, \ and\ \bibinfo
  {author} {\bibfnamefont {K.}~\bibnamefont {Yang}},\ }\href {\doibase
  10.1088/1475-7516/2012/10/053} {\bibfield  {journal} {\bibinfo  {journal}
  {J.~Cosmol.~A.~P.}\ }\textbf {\bibinfo {volume} {2012}},\ \bibinfo {pages}
  {053} (\bibinfo {year} {2012})}\BibitemShut {NoStop}%
\bibitem [{\citenamefont {Sotani}\ and\ \citenamefont
  {Miyamoto}(2015)}]{Sotani:2015a}%
  \BibitemOpen
  \bibfield  {author} {\bibinfo {author} {\bibfnamefont {H.}~\bibnamefont
  {Sotani}}\ and\ \bibinfo {author} {\bibfnamefont {U.}~\bibnamefont
  {Miyamoto}},\ }\href {\doibase 10.1103/PhysRevD.92.044052} {\bibfield
  {journal} {\bibinfo  {journal} {Phys.~Rev.~D.}\ }\textbf {\bibinfo {volume}
  {92}},\ \bibinfo {pages} {044052} (\bibinfo {year} {2015})}\BibitemShut
  {NoStop}%
\bibitem [{\citenamefont {Zhao}\ and\ \citenamefont {Xie}(2017)}]{Zhao:2017a}%
  \BibitemOpen
  \bibfield  {author} {\bibinfo {author} {\bibfnamefont {S.-S.}\ \bibnamefont
  {Zhao}}\ and\ \bibinfo {author} {\bibfnamefont {Y.}~\bibnamefont {Xie}},\
  }\href {\doibase 10.1016/j.physletb.2017.09.090} {\bibfield  {journal}
  {\bibinfo  {journal} {Phys.~Lett.~B.}\ }\textbf {\bibinfo {volume} {774}},\
  \bibinfo {pages} {357} (\bibinfo {year} {2017})}\BibitemShut {NoStop}%
\bibitem [{\citenamefont {{Chakraborty}}\ and\ \citenamefont
  {Soumitra}(2017)}]{Chak:2017a}%
  \BibitemOpen
  \bibfield  {author} {\bibinfo {author} {\bibfnamefont {S.}~\bibnamefont
  {{Chakraborty}}}\ and\ \bibinfo {author} {\bibfnamefont {S.}~\bibnamefont
  {Soumitra}},\ }\href {\doibase 10.1088/1475-7516/2017/07/045} {\bibfield
  {journal} {\bibinfo  {journal} {J.~Cosmol.~A.~P.}\ }\textbf {\bibinfo
  {volume} {07}},\ \bibinfo {pages} {045} (\bibinfo {year} {2017})}\BibitemShut
  {NoStop}%
\bibitem [{\citenamefont {{Jin}}\ \emph {et~al.}(2020)\citenamefont {{Jin}},
  \citenamefont {{Gao}},\ and\ \citenamefont {{Liu}}}]{Jin:2020a}%
  \BibitemOpen
  \bibfield  {author} {\bibinfo {author} {\bibfnamefont {X.-H.}\ \bibnamefont
  {{Jin}}}, \bibinfo {author} {\bibfnamefont {Y.-X.}\ \bibnamefont {{Gao}}}, \
  and\ \bibinfo {author} {\bibfnamefont {D.-J.}\ \bibnamefont {{Liu}}},\ }\href
  {\doibase 10.1142/S0218271820500650} {\bibfield  {journal} {\bibinfo
  {journal} {International Journal of Modern Physics D}\ }\textbf {\bibinfo
  {volume} {29}},\ \bibinfo {eid} {2050065} (\bibinfo {year}
  {2020})}\BibitemShut {NoStop}%
\bibitem [{\citenamefont {Bisnovatyi-Kogan}\ and\ \citenamefont
  {Tsupko}(2010)}]{Bin:2010a}%
  \BibitemOpen
  \bibfield  {author} {\bibinfo {author} {\bibfnamefont {G.~S.}\ \bibnamefont
  {Bisnovatyi-Kogan}}\ and\ \bibinfo {author} {\bibfnamefont {O.~Y.}\
  \bibnamefont {Tsupko}},\ }\href {\doibase 10.1111/j.1365-2966.2010.16290.x}
  {\bibfield  {journal} {\bibinfo  {journal} {Mon.~Not.~R.~Astron.~Soc.}\
  }\textbf {\bibinfo {volume} {404}},\ \bibinfo {pages} {1790} (\bibinfo {year}
  {2010})}\BibitemShut {NoStop}%
\bibitem [{\citenamefont {Tsupko}\ and\ \citenamefont
  {Bisnovatyi-Kogan}(2012)}]{Tsp:2011a}%
  \BibitemOpen
  \bibfield  {author} {\bibinfo {author} {\bibfnamefont {O.~Y.}\ \bibnamefont
  {Tsupko}}\ and\ \bibinfo {author} {\bibfnamefont {G.~S.}\ \bibnamefont
  {Bisnovatyi-Kogan}},\ }\href {\doibase 10.1134/S0202289312020120} {\bibfield
  {journal} {\bibinfo  {journal} {Gravit.~Cosmol.}\ }\textbf {\bibinfo {volume}
  {18}},\ \bibinfo {pages} {117} (\bibinfo {year} {2012})}\BibitemShut
  {NoStop}%
\bibitem [{\citenamefont {Tsupko}\ and\ \citenamefont
  {Bisnovatyi-Kogan}(2014)}]{Tsp:2014a}%
  \BibitemOpen
  \bibfield  {author} {\bibinfo {author} {\bibfnamefont {O.~Y.}\ \bibnamefont
  {Tsupko}}\ and\ \bibinfo {author} {\bibfnamefont {G.~S.}\ \bibnamefont
  {Bisnovatyi-Kogan}},\ }\href {\doibase 10.1134/S0202289314030153} {\bibfield
  {journal} {\bibinfo  {journal} {Gravit.~Cosmol.}\ }\textbf {\bibinfo {volume}
  {20}},\ \bibinfo {pages} {220} (\bibinfo {year} {2014})}\BibitemShut
  {NoStop}%
\bibitem [{\citenamefont {Tsupko}\ and\ \citenamefont
  {Bisnovatyi-Kogan}(2015)}]{Tsp:2015a}%
  \BibitemOpen
  \bibfield  {author} {\bibinfo {author} {\bibfnamefont {O.~Y.}\ \bibnamefont
  {Tsupko}}\ and\ \bibinfo {author} {\bibfnamefont {G.~S.}\ \bibnamefont
  {Bisnovatyi-Kogan}},\ }\href {\doibase 10.1134/S1063780X15070016} {\bibfield
  {journal} {\bibinfo  {journal} {Plasma Physics Reports}\ }\textbf {\bibinfo
  {volume} {41}},\ \bibinfo {pages} {562} (\bibinfo {year} {2015})}\BibitemShut
  {NoStop}%
\bibitem [{\citenamefont {Bisnovatyi-Kogan}\ and\ \citenamefont
  {Tsupko}(2017)}]{Bis:2017a}%
  \BibitemOpen
  \bibfield  {author} {\bibinfo {author} {\bibfnamefont {G.~S.}\ \bibnamefont
  {Bisnovatyi-Kogan}}\ and\ \bibinfo {author} {\bibfnamefont {O.~Y.}\
  \bibnamefont {Tsupko}},\ }\href {\doibase 10.3390/universe3030057} {\bibfield
   {journal} {\bibinfo  {journal} {Universe.}\ }\textbf {\bibinfo {volume}
  {3}},\ \bibinfo {pages} {1} (\bibinfo {year} {2017})}\BibitemShut {NoStop}%
\bibitem [{\citenamefont {Morozova}\ \emph {et~al.}(2013)\citenamefont
  {Morozova}, \citenamefont {Ahmedov},\ and\ \citenamefont
  {Tursunov}}]{Abu:2013a}%
  \BibitemOpen
  \bibfield  {author} {\bibinfo {author} {\bibfnamefont {V.}~\bibnamefont
  {Morozova}}, \bibinfo {author} {\bibfnamefont {B.}~\bibnamefont {Ahmedov}}, \
  and\ \bibinfo {author} {\bibfnamefont {A.}~\bibnamefont {Tursunov}},\ }\href
  {\doibase 10.1007/s10509-013-1458-6} {\bibfield  {journal} {\bibinfo
  {journal} {Astrophys.~Space.~Sci.}\ }\textbf {\bibinfo {volume} {346}},\
  \bibinfo {pages} {513} (\bibinfo {year} {2013})}\BibitemShut {NoStop}%
\bibitem [{\citenamefont {{Chakrabarty}}\ \emph {et~al.}(2018)\citenamefont
  {{Chakrabarty}}, \citenamefont {{Abdikamalov}}, \citenamefont
  {{Abdujabbarov}},\ and\ \citenamefont {{Bambi}}}]{Chak:2018a}%
  \BibitemOpen
  \bibfield  {author} {\bibinfo {author} {\bibfnamefont {H.}~\bibnamefont
  {{Chakrabarty}}}, \bibinfo {author} {\bibfnamefont {A.~B.}\ \bibnamefont
  {{Abdikamalov}}}, \bibinfo {author} {\bibfnamefont {A.~A.}\ \bibnamefont
  {{Abdujabbarov}}}, \ and\ \bibinfo {author} {\bibfnamefont {C.}~\bibnamefont
  {{Bambi}}},\ }\href {\doibase 10.1103/PhysRevD.98.024022} {\bibfield
  {journal} {\bibinfo  {journal} {Phys.~Rev.~D.}\ }\textbf {\bibinfo {volume}
  {98}},\ \bibinfo {pages} {024022} (\bibinfo {year} {2018})}\BibitemShut
  {NoStop}%
\bibitem [{\citenamefont {{Turimov}}\ \emph {et~al.}(2019)\citenamefont
  {{Turimov}}, \citenamefont {{Ahmedov}}, \citenamefont {{Abdujabbarov}},\ and\
  \citenamefont {{Bambi}}}]{Turi:2019a}%
  \BibitemOpen
  \bibfield  {author} {\bibinfo {author} {\bibfnamefont {B.}~\bibnamefont
  {{Turimov}}}, \bibinfo {author} {\bibfnamefont {B.}~\bibnamefont
  {{Ahmedov}}}, \bibinfo {author} {\bibfnamefont {A.}~\bibnamefont
  {{Abdujabbarov}}}, \ and\ \bibinfo {author} {\bibfnamefont {C.}~\bibnamefont
  {{Bambi}}},\ }\href {\doibase 10.1142/S0218271820400131} {\bibfield
  {journal} {\bibinfo  {journal} {Int.~J.~Mod.~Phys.~D.}\ }\textbf {\bibinfo
  {volume} {28}},\ \bibinfo {pages} {2040013} (\bibinfo {year}
  {2019})}\BibitemShut {NoStop}%
\bibitem [{\citenamefont {Hakimov}\ and\ \citenamefont
  {Atamurotov}(2016)}]{Far:2016a}%
  \BibitemOpen
  \bibfield  {author} {\bibinfo {author} {\bibfnamefont {A.}~\bibnamefont
  {Hakimov}}\ and\ \bibinfo {author} {\bibfnamefont {F.}~\bibnamefont
  {Atamurotov}},\ }\href {\doibase 10.1007/s10509-016-2702-7} {\bibfield
  {journal} {\bibinfo  {journal} {Astrophys.~Space.~Sci.}\ }\textbf {\bibinfo
  {volume} {361}},\ \bibinfo {pages} {112} (\bibinfo {year}
  {2016})}\BibitemShut {NoStop}%
\bibitem [{\citenamefont {Abdujabbarov}\ \emph
  {et~al.}(2017{\natexlab{a}})\citenamefont {Abdujabbarov}, \citenamefont
  {Toshmatov}, \citenamefont {Schee}, \citenamefont {Stuchl{\'\i}k},\ and\
  \citenamefont {Ahmedov}}]{Abu:2017aa}%
  \BibitemOpen
  \bibfield  {author} {\bibinfo {author} {\bibfnamefont {A.}~\bibnamefont
  {Abdujabbarov}}, \bibinfo {author} {\bibfnamefont {B.}~\bibnamefont
  {Toshmatov}}, \bibinfo {author} {\bibfnamefont {J.}~\bibnamefont {Schee}},
  \bibinfo {author} {\bibfnamefont {Z.}~\bibnamefont {Stuchl{\'\i}k}}, \ and\
  \bibinfo {author} {\bibfnamefont {B.}~\bibnamefont {Ahmedov}},\ }\href
  {\doibase 10.1142/S0218271817410115} {\bibfield  {journal} {\bibinfo
  {journal} {Int.~J.~Mod.~Phys.~D.}\ }\textbf {\bibinfo {volume} {26}},\
  \bibinfo {pages} {1741011} (\bibinfo {year}
  {2017}{\natexlab{a}})}\BibitemShut {NoStop}%
\bibitem [{\citenamefont {Abdujabbarov}\ \emph
  {et~al.}(2017{\natexlab{b}})\citenamefont {Abdujabbarov}, \citenamefont
  {Ahmedov}, \citenamefont {Dadhich},\ and\ \citenamefont
  {Atamurotov}}]{Abu:2017a}%
  \BibitemOpen
  \bibfield  {author} {\bibinfo {author} {\bibfnamefont {A.}~\bibnamefont
  {Abdujabbarov}}, \bibinfo {author} {\bibfnamefont {B.}~\bibnamefont
  {Ahmedov}}, \bibinfo {author} {\bibfnamefont {N.}~\bibnamefont {Dadhich}}, \
  and\ \bibinfo {author} {\bibfnamefont {F.}~\bibnamefont {Atamurotov}},\
  }\href {\doibase 10.1103/PhysRevD.96.084017} {\bibfield  {journal} {\bibinfo
  {journal} {Phys.~Rev.~D.}\ }\textbf {\bibinfo {volume} {96}},\ \bibinfo
  {pages} {084017} (\bibinfo {year} {2017}{\natexlab{b}})}\BibitemShut
  {NoStop}%
\bibitem [{\citenamefont {Benavides-Gallego}\ \emph {et~al.}(2018)\citenamefont
  {Benavides-Gallego}, \citenamefont {Abdujabbarov},\ and\ \citenamefont
  {Bambi}}]{Car:2018a}%
  \BibitemOpen
  \bibfield  {author} {\bibinfo {author} {\bibfnamefont {C.}~\bibnamefont
  {Benavides-Gallego}}, \bibinfo {author} {\bibfnamefont {A.}~\bibnamefont
  {Abdujabbarov}}, \ and\ \bibinfo {author} {\bibnamefont {Bambi}},\ }\href
  {\doibase 10.1140/epjc/s10052-018-6170-97} {\bibfield  {journal} {\bibinfo
  {journal} {Eur.~Phys.~J.~C.}\ }\textbf {\bibinfo {volume} {78}},\ \bibinfo
  {pages} {694} (\bibinfo {year} {2018})}\BibitemShut {NoStop}%
\bibitem [{\citenamefont {Babar}\ \emph {et~al.}(2020)\citenamefont {Babar},
  \citenamefont {Babar},\ and\ \citenamefont {Atamurotov}}]{Babar:2020a}%
  \BibitemOpen
  \bibfield  {author} {\bibinfo {author} {\bibfnamefont {G.~Z.}\ \bibnamefont
  {Babar}}, \bibinfo {author} {\bibfnamefont {A.~Z.}\ \bibnamefont {Babar}}, \
  and\ \bibinfo {author} {\bibfnamefont {F.}~\bibnamefont {Atamurotov}},\
  }\href {\doibase 10.1140/epjc/s10052-020-8346-3} {\bibfield  {journal}
  {\bibinfo  {journal} {Eur.~Phys.~J.~C.}\ }\textbf {\bibinfo {volume} {80}},\
  \bibinfo {pages} {761} (\bibinfo {year} {2020})}\BibitemShut {NoStop}%
\bibitem [{\citenamefont {{Chowdhuri}}\ and\ \citenamefont
  {{Bhattacharyya}}(2020)}]{Chow:2020a}%
  \BibitemOpen
  \bibfield  {author} {\bibinfo {author} {\bibfnamefont {A.}~\bibnamefont
  {{Chowdhuri}}}\ and\ \bibinfo {author} {\bibfnamefont {A.}~\bibnamefont
  {{Bhattacharyya}}},\ }\href {\doibase arXiv:astro-ph/2012.12914} {\bibfield
  {journal} {\bibinfo  {journal} {arXiv.org}\ } (\bibinfo {year} {2020}),\
  arXiv:astro-ph/2012.12914}\BibitemShut {NoStop}%
\bibitem [{\citenamefont {{Atamurotov}}\ \emph {et~al.}(2021)\citenamefont
  {{Atamurotov}}, \citenamefont {{Abdujabbarov}},\ and\ \citenamefont
  {{Rayimbaev}}}]{Far:2021a}%
  \BibitemOpen
  \bibfield  {author} {\bibinfo {author} {\bibfnamefont {F.}~\bibnamefont
  {{Atamurotov}}}, \bibinfo {author} {\bibfnamefont {A.}~\bibnamefont
  {{Abdujabbarov}}}, \ and\ \bibinfo {author} {\bibfnamefont {J.}~\bibnamefont
  {{Rayimbaev}}},\ }\href {\doibase 10.1140/epjc/s10052-021-08919-x} {\bibfield
   {journal} {\bibinfo  {journal} {Eur.~Phys.~J.~C.}\ }\textbf {\bibinfo
  {volume} {81}},\ \bibinfo {pages} {118} (\bibinfo {year} {2021})}\BibitemShut
  {NoStop}%
\bibitem [{\citenamefont {Glavan}\ and\ \citenamefont
  {Lin}(2020)}]{Glav:2020a}%
  \BibitemOpen
  \bibfield  {author} {\bibinfo {author} {\bibfnamefont {D.}~\bibnamefont
  {Glavan}}\ and\ \bibinfo {author} {\bibfnamefont {C.}~\bibnamefont {Lin}},\
  }\href {\doibase 10.1103/PhysRevLett.124.081301} {\bibfield  {journal}
  {\bibinfo  {journal} {Phys.~Rev.~Lett.}\ }\textbf {\bibinfo {volume} {124}},\
  \bibinfo {pages} {081301} (\bibinfo {year} {2020})}\BibitemShut {NoStop}%
\bibitem [{\citenamefont {Boulware}\ and\ \citenamefont
  {Deser}(1985)}]{Boul:1985a}%
  \BibitemOpen
  \bibfield  {author} {\bibinfo {author} {\bibfnamefont {D.~G.}\ \bibnamefont
  {Boulware}}\ and\ \bibinfo {author} {\bibfnamefont {S.}~\bibnamefont
  {Deser}},\ }\href {\doibase 10.1103/PhysRevLett.55.2656} {\bibfield
  {journal} {\bibinfo  {journal} {Phys.~Rev.~Lett.}\ }\textbf {\bibinfo
  {volume} {55}},\ \bibinfo {pages} {2656} (\bibinfo {year}
  {1985})}\BibitemShut {NoStop}%
\bibitem [{\citenamefont {ZWIEBACH}(1985)}]{Bart:1985b}%
  \BibitemOpen
  \bibfield  {author} {\bibinfo {author} {\bibfnamefont {B.}~\bibnamefont
  {ZWIEBACH}},\ }\href {\doibase 10.1016/0370-2693(85)91616-8} {\bibfield
  {journal} {\bibinfo  {journal} {Phys.~Lett.~B.}\ }\textbf {\bibinfo {volume}
  {156}},\ \bibinfo {pages} {315} (\bibinfo {year} {1985})}\BibitemShut
  {NoStop}%
\bibitem [{\citenamefont {Cai}(2002)}]{Cai:2002a}%
  \BibitemOpen
  \bibfield  {author} {\bibinfo {author} {\bibfnamefont {R.-G.}\ \bibnamefont
  {Cai}},\ }\href {\doibase 10.1103/PhysRevD.65.084014} {\bibfield  {journal}
  {\bibinfo  {journal} {Phys.~Rev.~D.}\ }\textbf {\bibinfo {volume} {65}},\
  \bibinfo {pages} {084014} (\bibinfo {year} {2002})}\BibitemShut {NoStop}%
\bibitem [{\citenamefont {Lovelock}(1971)}]{Dav:1971a}%
  \BibitemOpen
  \bibfield  {author} {\bibinfo {author} {\bibfnamefont {D.}~\bibnamefont
  {Lovelock}},\ }\href {\doibase 10.1063/1.1665613} {\bibfield  {journal}
  {\bibinfo  {journal} {J.~Math.~Phys.}\ }\textbf {\bibinfo {volume} {12}},\
  \bibinfo {pages} {498} (\bibinfo {year} {1971})}\BibitemShut {NoStop}%
\bibitem [{\citenamefont {G$\ddot{u}$rses}\ \emph
  {et~al.}(2020{\natexlab{a}})\citenamefont {G$\ddot{u}$rses}, \citenamefont
  {Sisman},\ and\ \citenamefont {Tekin}}]{Mg:2020a}%
  \BibitemOpen
  \bibfield  {author} {\bibinfo {author} {\bibfnamefont {M.}~\bibnamefont
  {G$\ddot{u}$rses}}, \bibinfo {author} {\bibfnamefont {T.~C.}\ \bibnamefont
  {Sisman}}, \ and\ \bibinfo {author} {\bibfnamefont {B.}~\bibnamefont
  {Tekin}},\ }\href {\doibase 10.1103/PhysRevLett.125.149001} {\bibfield
  {journal} {\bibinfo  {journal} {Phys.~ Rev.~Lett.}\ }\textbf {\bibinfo
  {volume} {125}},\ \bibinfo {pages} {149001} (\bibinfo {year}
  {2020}{\natexlab{a}})}\BibitemShut {NoStop}%
\bibitem [{\citenamefont {Arrechea}\ \emph
  {et~al.}(2020{\natexlab{a}})\citenamefont {Arrechea}, \citenamefont
  {Delhom},\ and\ \citenamefont {Jim$\acute{e}$nez-Cano}}]{Jul:2020a}%
  \BibitemOpen
  \bibfield  {author} {\bibinfo {author} {\bibfnamefont {J.}~\bibnamefont
  {Arrechea}}, \bibinfo {author} {\bibfnamefont {A.}~\bibnamefont {Delhom}}, \
  and\ \bibinfo {author} {\bibfnamefont {A.}~\bibnamefont
  {Jim$\acute{e}$nez-Cano}},\ }\href {\doibase 10.1103/PhysRevLett.125.149002}
  {\bibfield  {journal} {\bibinfo  {journal} {Phys.~ Rev.~Lett.}\ }\textbf
  {\bibinfo {volume} {125}},\ \bibinfo {pages} {149002} (\bibinfo {year}
  {2020}{\natexlab{a}})}\BibitemShut {NoStop}%
\bibitem [{\citenamefont {Arrechea}\ \emph {et~al.}(2021)\citenamefont
  {Arrechea}, \citenamefont {Delhom},\ and\ \citenamefont
  {Jim$\acute{e}$nez-Cano}}]{Jul:2020b}%
  \BibitemOpen
  \bibfield  {author} {\bibinfo {author} {\bibfnamefont {J.}~\bibnamefont
  {Arrechea}}, \bibinfo {author} {\bibfnamefont {A.}~\bibnamefont {Delhom}}, \
  and\ \bibinfo {author} {\bibfnamefont {A.}~\bibnamefont
  {Jim$\acute{e}$nez-Cano}},\ }\href {\doibase 10.1088/1674-1137/abc1d4}
  {\bibfield  {journal} {\bibinfo  {journal} {Chin.~Phys.~C}\ }\textbf
  {\bibinfo {volume} {45}},\ \bibinfo {pages} {013107} (\bibinfo {year}
  {2021})}\BibitemShut {NoStop}%
\bibitem [{\citenamefont {Malafarina}\ \emph {et~al.}(2020)\citenamefont
  {Malafarina}, \citenamefont {Toshmatov},\ and\ \citenamefont
  {Dadhich}}]{Daniel:2020a}%
  \BibitemOpen
  \bibfield  {author} {\bibinfo {author} {\bibfnamefont {D.}~\bibnamefont
  {Malafarina}}, \bibinfo {author} {\bibfnamefont {B.}~\bibnamefont
  {Toshmatov}}, \ and\ \bibinfo {author} {\bibfnamefont {N.}~\bibnamefont
  {Dadhich}},\ }\href {\doibase 10.1016/j.dark.2020.100598} {\bibfield
  {journal} {\bibinfo  {journal} {Physics of the Dark Universe}\ }\textbf
  {\bibinfo {volume} {30}},\ \bibinfo {pages} {100598} (\bibinfo {year}
  {2020})}\BibitemShut {NoStop}%
\bibitem [{\citenamefont {Zhang}\ \emph
  {et~al.}(2020{\natexlab{a}})\citenamefont {Zhang}, \citenamefont {Zhang},
  \citenamefont {Li},\ and\ \citenamefont {Guo}}]{Zhang:2020a}%
  \BibitemOpen
  \bibfield  {author} {\bibinfo {author} {\bibfnamefont {C.-Y.}\ \bibnamefont
  {Zhang}}, \bibinfo {author} {\bibfnamefont {S.-J.}\ \bibnamefont {Zhang}},
  \bibinfo {author} {\bibfnamefont {P.-C.}\ \bibnamefont {Li}}, \ and\ \bibinfo
  {author} {\bibfnamefont {M.}~\bibnamefont {Guo}},\ }\href {\doibase
  10.1007/JHEP08(2020)105} {\bibfield  {journal} {\bibinfo  {journal} {J.~High
  Energ.~Phys.}\ }\textbf {\bibinfo {volume} {2020}},\ \bibinfo {pages} {105}
  (\bibinfo {year} {2020}{\natexlab{a}})}\BibitemShut {NoStop}%
\bibitem [{\citenamefont {Aguilar-P$\acute{e}$rez}\ \emph
  {et~al.}(2019)\citenamefont {Aguilar-P$\acute{e}$rez}, \citenamefont {Cruzy},
  \citenamefont {Lepez},\ and\ \citenamefont {Moran-Riverax}}]{Gilb:2019a}%
  \BibitemOpen
  \bibfield  {author} {\bibinfo {author} {\bibfnamefont {G.}~\bibnamefont
  {Aguilar-P$\acute{e}$rez}}, \bibinfo {author} {\bibfnamefont
  {M.}~\bibnamefont {Cruzy}}, \bibinfo {author} {\bibfnamefont
  {S.}~\bibnamefont {Lepez}}, \ and\ \bibinfo {author} {\bibfnamefont
  {I.}~\bibnamefont {Moran-Riverax}},\ }\href {\doibase
  arXiv:astro-ph/1907.06168} {\bibfield  {journal} {\bibinfo  {journal}
  {arXiv.org}\ } (\bibinfo {year} {2019}),\
  arXiv:astro-ph/1907.06168}\BibitemShut {NoStop}%
\bibitem [{\citenamefont {Zhang}\ \emph
  {et~al.}(2020{\natexlab{b}})\citenamefont {Zhang}, \citenamefont {Wei},\ and\
  \citenamefont {Liu}}]{Yupeng:2020a}%
  \BibitemOpen
  \bibfield  {author} {\bibinfo {author} {\bibfnamefont {Y.-P.}\ \bibnamefont
  {Zhang}}, \bibinfo {author} {\bibfnamefont {S.-W.}\ \bibnamefont {Wei}}, \
  and\ \bibinfo {author} {\bibfnamefont {Y.-X.}\ \bibnamefont {Liu}},\ }\href
  {\doibase arXiv:astro-ph/2003.10960} {\bibfield  {journal} {\bibinfo
  {journal} {arXiv.org}\ } (\bibinfo {year} {2020}{\natexlab{b}}),\
  arXiv:astro-ph/2003.10960}\BibitemShut {NoStop}%
\bibitem [{\citenamefont {Guo}\ and\ \citenamefont {Li}(2020)}]{Ming:2020a}%
  \BibitemOpen
  \bibfield  {author} {\bibinfo {author} {\bibfnamefont {M.}~\bibnamefont
  {Guo}}\ and\ \bibinfo {author} {\bibfnamefont {P.-C.}\ \bibnamefont {Li}},\
  }\href {\doibase 10.1140/epjc/s10052-020-8164-7} {\bibfield  {journal}
  {\bibinfo  {journal} {Eur.~Phys.~J.~C.}\ }\textbf {\bibinfo {volume} {80}},\
  \bibinfo {pages} {588} (\bibinfo {year} {2020})}\BibitemShut {NoStop}%
\bibitem [{\citenamefont {Kumar}\ and\ \citenamefont
  {Ghosh}(2020)}]{Rahul:2020a}%
  \BibitemOpen
  \bibfield  {author} {\bibinfo {author} {\bibfnamefont {R.}~\bibnamefont
  {Kumar}}\ and\ \bibinfo {author} {\bibfnamefont {S.~G.}\ \bibnamefont
  {Ghosh}},\ }\href {\doibase 10.1088/1475-7516/2020/07/053} {\bibfield
  {journal} {\bibinfo  {journal} {J.~Cosmol.~A.~P}\ }\textbf {\bibinfo {volume}
  {2020}},\ \bibinfo {pages} {053} (\bibinfo {year} {2020})}\BibitemShut
  {NoStop}%
\bibitem [{\citenamefont {Mishra}(2020)}]{Akash:2020a}%
  \BibitemOpen
  \bibfield  {author} {\bibinfo {author} {\bibfnamefont {A.~K.}\ \bibnamefont
  {Mishra}},\ }\href {\doibase arXiv:astro-ph/2004.01243} {\bibfield  {journal}
  {\bibinfo  {journal} {arXiv.org}\ } (\bibinfo {year} {2020}),\
  arXiv:astro-ph/2004.01243}\BibitemShut {NoStop}%
\bibitem [{\citenamefont {Churilova}(2020)}]{Churi:2020a}%
  \BibitemOpen
  \bibfield  {author} {\bibinfo {author} {\bibfnamefont {M.~S.}\ \bibnamefont
  {Churilova}},\ }\href {\doibase arXiv:astro-ph/2004.00513} {\bibfield
  {journal} {\bibinfo  {journal} {arXiv.org}\ } (\bibinfo {year} {2020}),\
  arXiv:astro-ph/2004.00513}\BibitemShut {NoStop}%
\bibitem [{\citenamefont {Arag$\acute{o}$n}\ \emph {et~al.}(2020)\citenamefont
  {Arag$\acute{o}$n}, \citenamefont {B$\acute{e}$car}, \citenamefont
  {Gonz$\acute{a}$lez},\ and\ \citenamefont {V$\acute{a}$squez}}]{Almen:2020a}%
  \BibitemOpen
  \bibfield  {author} {\bibinfo {author} {\bibfnamefont {A.}~\bibnamefont
  {Arag$\acute{o}$n}}, \bibinfo {author} {\bibfnamefont {R.}~\bibnamefont
  {B$\acute{e}$car}}, \bibinfo {author} {\bibfnamefont {P.~A.}\ \bibnamefont
  {Gonz$\acute{a}$lez}}, \ and\ \bibinfo {author} {\bibfnamefont
  {Y.}~\bibnamefont {V$\acute{a}$squez}},\ }\href {\doibase
  10.1140/epjc/s10052-020-8298-7} {\bibfield  {journal} {\bibinfo  {journal}
  {Eur.~Phys.~J.~C.}\ }\textbf {\bibinfo {volume} {80}},\ \bibinfo {pages}
  {773} (\bibinfo {year} {2020})}\BibitemShut {NoStop}%
\bibitem [{\citenamefont {Islam}\ \emph {et~al.}(2020)\citenamefont {Islam},
  \citenamefont {Kumara},\ and\ \citenamefont {Ghosha}}]{Islam:2020a}%
  \BibitemOpen
  \bibfield  {author} {\bibinfo {author} {\bibfnamefont {S.~U.}\ \bibnamefont
  {Islam}}, \bibinfo {author} {\bibfnamefont {R.}~\bibnamefont {Kumara}}, \
  and\ \bibinfo {author} {\bibfnamefont {S.~G.}\ \bibnamefont {Ghosha}},\
  }\href {\doibase 10.1088/1475-7516/2020/09/030} {\bibfield  {journal}
  {\bibinfo  {journal} {J.~Cosmol.~A.~P}\ }\textbf {\bibinfo {volume} {2020}},\
  \bibinfo {pages} {030} (\bibinfo {year} {2020})}\BibitemShut {NoStop}%
\bibitem [{\citenamefont {Abdujabbarov}\ \emph {et~al.}(2015)\citenamefont
  {Abdujabbarov}, \citenamefont {Atamurotov}, \citenamefont {Dadhich},
  \citenamefont {Ahmedov},\ and\ \citenamefont
  {Stuchl$\acute{i}$k}}]{Abu:2015d}%
  \BibitemOpen
  \bibfield  {author} {\bibinfo {author} {\bibfnamefont {A.}~\bibnamefont
  {Abdujabbarov}}, \bibinfo {author} {\bibfnamefont {F.}~\bibnamefont
  {Atamurotov}}, \bibinfo {author} {\bibfnamefont {N.}~\bibnamefont {Dadhich}},
  \bibinfo {author} {\bibfnamefont {B.}~\bibnamefont {Ahmedov}}, \ and\
  \bibinfo {author} {\bibfnamefont {Z.}~\bibnamefont {Stuchl$\acute{i}$k}},\
  }\href {\doibase 10.1140/epjc/s10052-015-3604-5} {\bibfield  {journal}
  {\bibinfo  {journal} {Eur.~Phys.~J.~C.}\ }\textbf {\bibinfo {volume} {75}},\
  \bibinfo {pages} {399} (\bibinfo {year} {2015})}\BibitemShut {NoStop}%
\bibitem [{\citenamefont {Hegde}\ \emph {et~al.}(2020)\citenamefont {Hegde},
  \citenamefont {Kumara~A.}, \citenamefont {Rizwan~C.L.}, \citenamefont
  {M.~Ajith},\ and\ \citenamefont {Sabir~Ali}}]{Hedge:2020a}%
  \BibitemOpen
  \bibfield  {author} {\bibinfo {author} {\bibfnamefont {K.}~\bibnamefont
  {Hegde}}, \bibinfo {author} {\bibfnamefont {N.}~\bibnamefont {Kumara~A.}},
  \bibinfo {author} {\bibfnamefont {A.}~\bibnamefont {Rizwan~C.L.}}, \bibinfo
  {author} {\bibfnamefont {K.}~\bibnamefont {M.~Ajith}}, \ and\ \bibinfo
  {author} {\bibfnamefont {M.}~\bibnamefont {Sabir~Ali}},\ }\href {\doibase
  arXiv:astro-ph/2003.08778} {\bibfield  {journal} {\bibinfo  {journal}
  {arXiv.org}\ } (\bibinfo {year} {2020}),\
  arXiv:astro-ph/2003.08778}\BibitemShut {NoStop}%
\bibitem [{\citenamefont {Mansoori}(2020)}]{Mansr:2020a}%
  \BibitemOpen
  \bibfield  {author} {\bibinfo {author} {\bibfnamefont {S.~A.~H.}\
  \bibnamefont {Mansoori}},\ }\href {\doibase arXiv:astro-ph/2003.13382}
  {\bibfield  {journal} {\bibinfo  {journal} {arXiv.org}\ } (\bibinfo {year}
  {2020}),\ arXiv:astro-ph/2003.13382}\BibitemShut {NoStop}%
\bibitem [{\citenamefont {Kumara~A.}\ \emph {et~al.}(2020)\citenamefont
  {Kumara~A.}, \citenamefont {Rizwan~C.L.}, \citenamefont {Hegde},
  \citenamefont {M.~Ajith},\ and\ \citenamefont {Sabir~Ali}}]{Kumara:2020a}%
  \BibitemOpen
  \bibfield  {author} {\bibinfo {author} {\bibfnamefont {N.}~\bibnamefont
  {Kumara~A.}}, \bibinfo {author} {\bibfnamefont {A.}~\bibnamefont
  {Rizwan~C.L.}}, \bibinfo {author} {\bibfnamefont {K.}~\bibnamefont {Hegde}},
  \bibinfo {author} {\bibfnamefont {K.}~\bibnamefont {M.~Ajith}}, \ and\
  \bibinfo {author} {\bibfnamefont {M.}~\bibnamefont {Sabir~Ali}},\ }\href
  {\doibase arXiv:astro-ph/2004.04521} {\bibfield  {journal} {\bibinfo
  {journal} {arXiv.org}\ } (\bibinfo {year} {2020}),\
  arXiv:astro-ph/2004.04521}\BibitemShut {NoStop}%
\bibitem [{\citenamefont {Li}\ \emph {et~al.}(2020)\citenamefont {Li},
  \citenamefont {Wu},\ and\ \citenamefont {Yu}}]{Li:2020a}%
  \BibitemOpen
  \bibfield  {author} {\bibinfo {author} {\bibfnamefont {S.-L.}\ \bibnamefont
  {Li}}, \bibinfo {author} {\bibfnamefont {P.}~\bibnamefont {Wu}}, \ and\
  \bibinfo {author} {\bibfnamefont {H.}~\bibnamefont {Yu}},\ }\href {\doibase
  arXiv:astro-ph/2004.02080} {\bibfield  {journal} {\bibinfo  {journal}
  {arXiv.org}\ } (\bibinfo {year} {2020}),\
  arXiv:astro-ph/2004.02080}\BibitemShut {NoStop}%
\bibitem [{\citenamefont {Wei}\ and\ \citenamefont
  {Liu}(2020{\natexlab{a}})}]{Wei:2020b}%
  \BibitemOpen
  \bibfield  {author} {\bibinfo {author} {\bibfnamefont {S.-W.}\ \bibnamefont
  {Wei}}\ and\ \bibinfo {author} {\bibfnamefont {Y.-X.}\ \bibnamefont {Liu}},\
  }\href {\doibase 10.1103/PhysRevD.101.104018} {\bibfield  {journal} {\bibinfo
   {journal} {Phys.~Rev.~D.}\ }\textbf {\bibinfo {volume} {101}},\ \bibinfo
  {pages} {104018} (\bibinfo {year} {2020}{\natexlab{a}})}\BibitemShut
  {NoStop}%
\bibitem [{\citenamefont {Veer~Singh}\ and\ \citenamefont
  {Siwach}(2020)}]{Dhrm:2020a}%
  \BibitemOpen
  \bibfield  {author} {\bibinfo {author} {\bibfnamefont {D.}~\bibnamefont
  {Veer~Singh}}\ and\ \bibinfo {author} {\bibfnamefont {S.}~\bibnamefont
  {Siwach}},\ }\href@noop {} {\bibfield  {journal} {\bibinfo  {journal}
  {Phys.~Lett.~B.}\ }\textbf {\bibinfo {volume} {808}} (\bibinfo {year}
  {2020})}\BibitemShut {NoStop}%
\bibitem [{\citenamefont {Wei}\ and\ \citenamefont
  {Liu}(2020{\natexlab{b}})}]{Wei:2020c}%
  \BibitemOpen
  \bibfield  {author} {\bibinfo {author} {\bibfnamefont {S.-W.}\ \bibnamefont
  {Wei}}\ and\ \bibinfo {author} {\bibfnamefont {Y.-X.}\ \bibnamefont {Liu}},\
  }\href {\doibase arXiv:astro-ph/2003.07769} {\bibfield  {journal} {\bibinfo
  {journal} {arXiv.org}\ } (\bibinfo {year} {2020}{\natexlab{b}}),\
  arXiv:astro-ph/2003.07769}\BibitemShut {NoStop}%
\bibitem [{\citenamefont {Heydari-Fard}\ \emph {et~al.}(2020)\citenamefont
  {Heydari-Fard}, \citenamefont {Heydari-Fard},\ and\ \citenamefont
  {Sepangi}}]{Moha:2020a}%
  \BibitemOpen
  \bibfield  {author} {\bibinfo {author} {\bibfnamefont {M.}~\bibnamefont
  {Heydari-Fard}}, \bibinfo {author} {\bibfnamefont {M.}~\bibnamefont
  {Heydari-Fard}}, \ and\ \bibinfo {author} {\bibfnamefont {H.~R.}\
  \bibnamefont {Sepangi}},\ }\href {\doibase arXiv:astro-ph/2004.02140}
  {\bibfield  {journal} {\bibinfo  {journal} {arXiv.org}\ } (\bibinfo {year}
  {2020}),\ arXiv:astro-ph/2004.02140}\BibitemShut {NoStop}%
\bibitem [{\citenamefont {G$\ddot{u}$rses}\ \emph
  {et~al.}(2020{\natexlab{b}})\citenamefont {G$\ddot{u}$rses}, \citenamefont
  {Sisman},\ and\ \citenamefont {Tekin}}]{Met:2020a}%
  \BibitemOpen
  \bibfield  {author} {\bibinfo {author} {\bibfnamefont {M.}~\bibnamefont
  {G$\ddot{u}$rses}}, \bibinfo {author} {\bibfnamefont {T.~C.}\ \bibnamefont
  {Sisman}}, \ and\ \bibinfo {author} {\bibfnamefont {B.}~\bibnamefont
  {Tekin}},\ }\href {\doibase 0.1140/epjc/s10052-020-8200-7} {\bibfield
  {journal} {\bibinfo  {journal} {Eur.~Phys.~J.~C.}\ }\textbf {\bibinfo
  {volume} {80}},\ \bibinfo {pages} {647} (\bibinfo {year}
  {2020}{\natexlab{b}})}\BibitemShut {NoStop}%
\bibitem [{\citenamefont {A.~Hennigar}\ \emph {et~al.}(2020)\citenamefont
  {A.~Hennigar}, \citenamefont {Kubiz$\check{n}\acute{a}$k}, \citenamefont
  {Mann},\ and\ \citenamefont {Pollack}}]{Robie:2020a}%
  \BibitemOpen
  \bibfield  {author} {\bibinfo {author} {\bibfnamefont {R.}~\bibnamefont
  {A.~Hennigar}}, \bibinfo {author} {\bibfnamefont {D.}~\bibnamefont
  {Kubiz$\check{n}\acute{a}$k}}, \bibinfo {author} {\bibfnamefont {R.~B.}\
  \bibnamefont {Mann}}, \ and\ \bibinfo {author} {\bibfnamefont
  {C.}~\bibnamefont {Pollack}},\ }\href {\doibase 10.1007/JHEP07(2020)027}
  {\bibfield  {journal} {\bibinfo  {journal} {J.~High Energ.~Phys.}\ }\textbf
  {\bibinfo {volume} {2020}},\ \bibinfo {pages} {27} (\bibinfo {year}
  {2020})}\BibitemShut {NoStop}%
\bibitem [{\citenamefont {Arrechea}\ \emph
  {et~al.}(2020{\natexlab{b}})\citenamefont {Arrechea}, \citenamefont
  {Delhom},\ and\ \citenamefont {Jim$\acute{e}$nez-Cano}}]{Juli:2020a}%
  \BibitemOpen
  \bibfield  {author} {\bibinfo {author} {\bibfnamefont {J.}~\bibnamefont
  {Arrechea}}, \bibinfo {author} {\bibfnamefont {A.}~\bibnamefont {Delhom}}, \
  and\ \bibinfo {author} {\bibfnamefont {A.}~\bibnamefont
  {Jim$\acute{e}$nez-Cano}},\ }\href {\doibase arXiv:astro-ph/2004.12998}
  {\bibfield  {journal} {\bibinfo  {journal} {arXiv.org}\ } (\bibinfo {year}
  {2020}{\natexlab{b}}),\ arXiv:astro-ph/2004.12998}\BibitemShut {NoStop}%
\bibitem [{\citenamefont {X.~Tian}\ and\ \citenamefont
  {Zhu}(2020)}]{Tian:2020a}%
  \BibitemOpen
  \bibfield  {author} {\bibinfo {author} {\bibfnamefont {S.}~\bibnamefont
  {X.~Tian}}\ and\ \bibinfo {author} {\bibfnamefont {Z.-H.}\ \bibnamefont
  {Zhu}},\ }\href {\doibase arXiv:astro-ph/2004.09954} {\bibfield  {journal}
  {\bibinfo  {journal} {arXiv.org}\ } (\bibinfo {year} {2020}),\
  arXiv:astro-ph/2004.09954}\BibitemShut {NoStop}%
\bibitem [{\citenamefont {Bonifacio}\ \emph {et~al.}(2020)\citenamefont
  {Bonifacio}, \citenamefont {Hinterbichler},\ and\ \citenamefont
  {A.~Johnson}}]{Jame:2020a}%
  \BibitemOpen
  \bibfield  {author} {\bibinfo {author} {\bibfnamefont {J.}~\bibnamefont
  {Bonifacio}}, \bibinfo {author} {\bibfnamefont {K.}~\bibnamefont
  {Hinterbichler}}, \ and\ \bibinfo {author} {\bibfnamefont {L.}~\bibnamefont
  {A.~Johnson}},\ }\href {\doibase 10.1103/PhysRevD.102.024029} {\bibfield
  {journal} {\bibinfo  {journal} {Phys.~Rev.~D.}\ }\textbf {\bibinfo {volume}
  {102}},\ \bibinfo {pages} {024029} (\bibinfo {year} {2020})}\BibitemShut
  {NoStop}%
\bibitem [{\citenamefont {L$\ddot{u}$~YiPang}(2020)}]{Yipng:2020a}%
  \BibitemOpen
  \bibfield  {author} {\bibinfo {author} {\bibfnamefont {H.}~\bibnamefont
  {L$\ddot{u}$~YiPang}},\ }\href@noop {} {\bibfield  {journal} {\bibinfo
  {journal} {Phys.~Lett.~B.}\ }\textbf {\bibinfo {volume} {809}} (\bibinfo
  {year} {2020})}\BibitemShut {NoStop}%
\bibitem [{\citenamefont {Kobayashi}(2020)}]{Tsut:2020a}%
  \BibitemOpen
  \bibfield  {author} {\bibinfo {author} {\bibfnamefont {T.}~\bibnamefont
  {Kobayashi}},\ }\href {\doibase 10.1088/1475-7516/2020/07/013} {\bibfield
  {journal} {\bibinfo  {journal} {J.~Cosmol.~A.~P}\ }\textbf {\bibinfo {volume}
  {2020}},\ \bibinfo {pages} {013} (\bibinfo {year} {2020})}\BibitemShut
  {NoStop}%
\bibitem [{\citenamefont {Fernandes}\ \emph {et~al.}(2020)\citenamefont
  {Fernandes}, \citenamefont {Carrilho}, \citenamefont {Clifton},\ and\
  \citenamefont {Mulryne}}]{Fern:2020a}%
  \BibitemOpen
  \bibfield  {author} {\bibinfo {author} {\bibfnamefont {P.~G.~S.}\
  \bibnamefont {Fernandes}}, \bibinfo {author} {\bibfnamefont {P.}~\bibnamefont
  {Carrilho}}, \bibinfo {author} {\bibfnamefont {T.}~\bibnamefont {Clifton}}, \
  and\ \bibinfo {author} {\bibfnamefont {D.~J.}\ \bibnamefont {Mulryne}},\
  }\href {\doibase 10.1103/PhysRevD.102.024025} {\bibfield  {journal} {\bibinfo
   {journal} {Phys.~Rev.~D.}\ }\textbf {\bibinfo {volume} {102}},\ \bibinfo
  {pages} {024025} (\bibinfo {year} {2020})}\BibitemShut {NoStop}%
\bibitem [{\citenamefont {Abdujabbarov}\ \emph {et~al.}(2020)\citenamefont
  {Abdujabbarov}, \citenamefont {Rayimbaev}, \citenamefont {Turimov},\ and\
  \citenamefont {Atamurotov}}]{Abu:2020f}%
  \BibitemOpen
  \bibfield  {author} {\bibinfo {author} {\bibfnamefont {A.}~\bibnamefont
  {Abdujabbarov}}, \bibinfo {author} {\bibfnamefont {J.}~\bibnamefont
  {Rayimbaev}}, \bibinfo {author} {\bibfnamefont {B.}~\bibnamefont {Turimov}},
  \ and\ \bibinfo {author} {\bibfnamefont {F.}~\bibnamefont {Atamurotov}},\
  }\href {\doibase 10.1016/j.dark.2020.100715} {\bibfield  {journal} {\bibinfo
  {journal} {Physics of the Dark Universe}\ }\textbf {\bibinfo {volume} {30}},\
  \bibinfo {pages} {100715} (\bibinfo {year} {2020})}\BibitemShut {NoStop}%
\bibitem [{\citenamefont {{Shaymatov}}\ \emph {et~al.}(2020)\citenamefont
  {{Shaymatov}}, \citenamefont {{Vrba}}, \citenamefont {{Malafarina}},
  \citenamefont {{Ahmedov}},\ and\ \citenamefont
  {{Stuchl{\'\i}k}}}]{Shay:2020a}%
  \BibitemOpen
  \bibfield  {author} {\bibinfo {author} {\bibfnamefont {S.}~\bibnamefont
  {{Shaymatov}}}, \bibinfo {author} {\bibfnamefont {J.}~\bibnamefont {{Vrba}}},
  \bibinfo {author} {\bibfnamefont {D.}~\bibnamefont {{Malafarina}}}, \bibinfo
  {author} {\bibfnamefont {B.}~\bibnamefont {{Ahmedov}}}, \ and\ \bibinfo
  {author} {\bibfnamefont {Z.}~\bibnamefont {{Stuchl{\'\i}k}}},\ }\href
  {\doibase 10.1016/j.dark.2020.100648} {\bibfield  {journal} {\bibinfo
  {journal} {Physics of the Dark Universe}\ }\textbf {\bibinfo {volume} {30}},\
  \bibinfo {pages} {100648} (\bibinfo {year} {2020})}\BibitemShut {NoStop}%
\bibitem [{\citenamefont {{Mahapatra}}(2020)}]{Maha:2020a}%
  \BibitemOpen
  \bibfield  {author} {\bibinfo {author} {\bibfnamefont {S.}~\bibnamefont
  {{Mahapatra}}},\ }\href {\doibase 10.1140/epjc/s10052-020-08568-6} {\bibfield
   {journal} {\bibinfo  {journal} {Eur.~Phys.~J.~C.}\ }\textbf {\bibinfo
  {volume} {80}},\ \bibinfo {pages} {992} (\bibinfo {year} {2020})}\BibitemShut
  {NoStop}%
\bibitem [{\citenamefont {{Feng}}\ \emph {et~al.}(2021)\citenamefont {{Feng}},
  \citenamefont {{Gu}},\ and\ \citenamefont {{Shu}}}]{Feng:2021a}%
  \BibitemOpen
  \bibfield  {author} {\bibinfo {author} {\bibfnamefont {J.-X.}\ \bibnamefont
  {{Feng}}}, \bibinfo {author} {\bibfnamefont {B.-M.}\ \bibnamefont {{Gu}}}, \
  and\ \bibinfo {author} {\bibfnamefont {F.-W.}\ \bibnamefont {{Shu}}},\ }\href
  {\doibase 10.1103/PhysRevD.103.064002} {\bibfield  {journal} {\bibinfo
  {journal} {Phys. Rev. D}\ }\textbf {\bibinfo {volume} {103}},\ \bibinfo {eid}
  {064002} (\bibinfo {year} {2021})}\BibitemShut {NoStop}%
\bibitem [{\citenamefont {{Mi{\v{s}}kovi{\'c}}}\ and\ \citenamefont
  {{Olea}}(2009)}]{Olivera:2009a}%
  \BibitemOpen
  \bibfield  {author} {\bibinfo {author} {\bibfnamefont {O.}~\bibnamefont
  {{Mi{\v{s}}kovi{\'c}}}}\ and\ \bibinfo {author} {\bibfnamefont
  {R.}~\bibnamefont {{Olea}}},\ }\href {\doibase 10.1103/PhysRevD.79.124020}
  {\bibfield  {journal} {\bibinfo  {journal} {Phys. Rev. D}\ }\textbf {\bibinfo
  {volume} {79}},\ \bibinfo {eid} {124020} (\bibinfo {year}
  {2009})}\BibitemShut {NoStop}%
\bibitem [{\citenamefont {Chandrasekhar}()}]{Chnd:1939a}%
  \BibitemOpen
  \bibfield  {author} {\bibinfo {author} {\bibfnamefont {S.}~\bibnamefont
  {Chandrasekhar}},\ }\href@noop {} {\bibinfo  {journal} {An introduction to
  the study of stellar structure (Dover Publications, New Haven, USA, 1939).
  Revised edition 1958}\ }\BibitemShut {NoStop}%
\bibitem [{\citenamefont {Binney}\ and\ \citenamefont
  {Tremaine}()}]{Jbin:1987a}%
  \BibitemOpen
\bibfield  {journal} {  }\bibfield  {author} {\bibinfo {author} {\bibfnamefont
  {J.}~\bibnamefont {Binney}}\ and\ \bibinfo {author} {\bibfnamefont
  {S.}~\bibnamefont {Tremaine}},\ }\href@noop {} {\bibinfo  {journal} {Galactic
  dynamics. Princeton, NJ, Princeton University Press, 1987, 747 p. (1987)}\
  }\BibitemShut {NoStop}%
\bibitem [{\citenamefont {Hinshaw}\ and\ \citenamefont
  {Krauss}(1987)}]{Hind:1987a}%
  \BibitemOpen
\bibfield  {journal} {  }\bibfield  {author} {\bibinfo {author} {\bibfnamefont
  {G.}~\bibnamefont {Hinshaw}}\ and\ \bibinfo {author} {\bibfnamefont {L.~M.}\
  \bibnamefont {Krauss}},\ }\href {\doibase 10.1086/165564} {\bibfield
  {journal} {\bibinfo  {journal} {Ap.~J.}\ }\textbf {\bibinfo {volume} {320}},\
  \bibinfo {pages} {468} (\bibinfo {year} {1987})}\BibitemShut {NoStop}%
\bibitem [{\citenamefont {Wu}(1996)}]{Wu:1996a}%
  \BibitemOpen
  \bibfield  {author} {\bibinfo {author} {\bibfnamefont {X.-P.}\ \bibnamefont
  {Wu}},\ }\href {\doibase arXiv:astro-ph/9512110} {\bibfield  {journal}
  {\bibinfo  {journal} {Fund.~Cosmic.~Phys.}\ }\textbf {\bibinfo {volume}
  {17}},\ \bibinfo {pages} {1} (\bibinfo {year} {1996})}\BibitemShut {NoStop}%
\end{thebibliography}%
\end{document}